\documentclass[preprint,12pt]{elsarticle}

\usepackage{amssymb,amsmath,amsfonts}
\usepackage{algorithmic}
\usepackage{mathdots}
\usepackage{graphicx}
\usepackage{textcomp}
\usepackage{xcolor}
\usepackage{lipsum}
\usepackage{floatrow}
\usepackage{subfig}
\usepackage{relsize}
\usepackage[hypcap=false]{caption}
\usepackage[figurename=Fig.]{caption}
\usepackage{longtable}
\usepackage{hyperref}
\usepackage{cleveref}
\usepackage{mathrsfs}
\usepackage{tabularray}
\usepackage{floatrow}
\usepackage{subcaption}
\usepackage{multirow} 
\usepackage{float}
\usepackage{comment}

\DeclareFloatFont{tiny}{\scriptsize}
\crefname{equation}{Eq}{Equations} 
\crefname{figure}{Fig}{Figs.}
\crefrangelabelformat{equation}{(#3#1#4--#5#2#6)}
\usepackage[a4paper, total={6.5in,10in}]{geometry}
\def\BibTeX{{\rm B\kern-.05em{\sc i\kern-.025em b}\kern-.08em
    T\kern-.1667em\lower.7ex\hbox{E}\kern-.125emX}}
\makeatletter
\def\ps@pprintTitle{%
  \let\@oddhead\@empty
  \let\@evenhead\@empty
  \def\@oddfoot{\reset@font\hfil\thepage\hfil}
  \let\@evenfoot\@oddfoot
}
\makeatother

\begin{document}

\begin{frontmatter}


\title{Comprehensive Dynamic Modeling and Constraint-Aware Air Supply Control for Localized Water Management in Automotive Polymer Electrolyte Membrane Fuel Cells}

\tnotetext[t1]{\textit{Submitted to Applied Energy}}

\author[add1]{Mostafaali Ayubirad}
\ead{Mostafa-Ali.Ayubirad@uvm.edu}

\author[add2]{Zeng Qiu}
\ead{cqiu1@ford.com}

\author[add2]{Hao Wang}
\ead{hwang210@ford.com}

\author[add2]{Chris Weinkauf}
\ead{cweinkau@ford.com}

\author[add2]{Michiel Van Nieuwstadt}
\ead{mvannie1@ford.com}

\author[add1]{Hamid R Ossareh}
\ead{Hamid.Ossareh@uvm.edu}

\affiliation[add1]{organization={Department of Electrical and Biomedical Engineering, University of Vermont},
          city={Burlington},
          postcode={05405}, 
          state={VT},
          country={USA}}

\affiliation[add2]{organization={Research \& Advanced Engineering, Ford Motor Company},
          city={Dearborn},
          postcode={48124}, 
          state={MI},
          country={USA}}

\begin{abstract}
In this paper, a predictive constraint-aware control scheme is formulated within the  Command Governor (CG) framework for localized hydration management of a proton exchange membrane (PEM) fuel cell system. First, a comprehensive nonlinear dynamic model of the fuel cell system is presented which includes a pseudo 2-dimensional (P2D) model of the stack, reactant supply and cooling  subsystems. The model captures the  couplings among the various subsystems and serves as the basis for designing output feedback controllers to track the optimal set-points of the air supply and cooling systems for power optimization. The closed-loop nonlinear model is then used to analyze the dynamic behavior of membrane hydration near the anode inlet, the driest region of the membrane in a counter-flow configuration, under various operating conditions. A reduced-order linearized model is then derived to approximate hydration behavior with sufficient fidelity for constraint enforcement. This model is used within the CG framework to adjust the air supply set-points when necessary to prevent membrane dry-out.  The effectiveness of the proposed approach in maintaining local membrane hydration while closely tracking the requested net power is demonstrated through realistic drive-cycle simulations.

\end{abstract}



\end{frontmatter}


\section{Introduction }
\label{sec:Introduction}
\begin{figure}
\centering
\vspace*{-2.0cm}  
\hspace*{-3.0cm}
\includegraphics[width=1.27\textwidth]{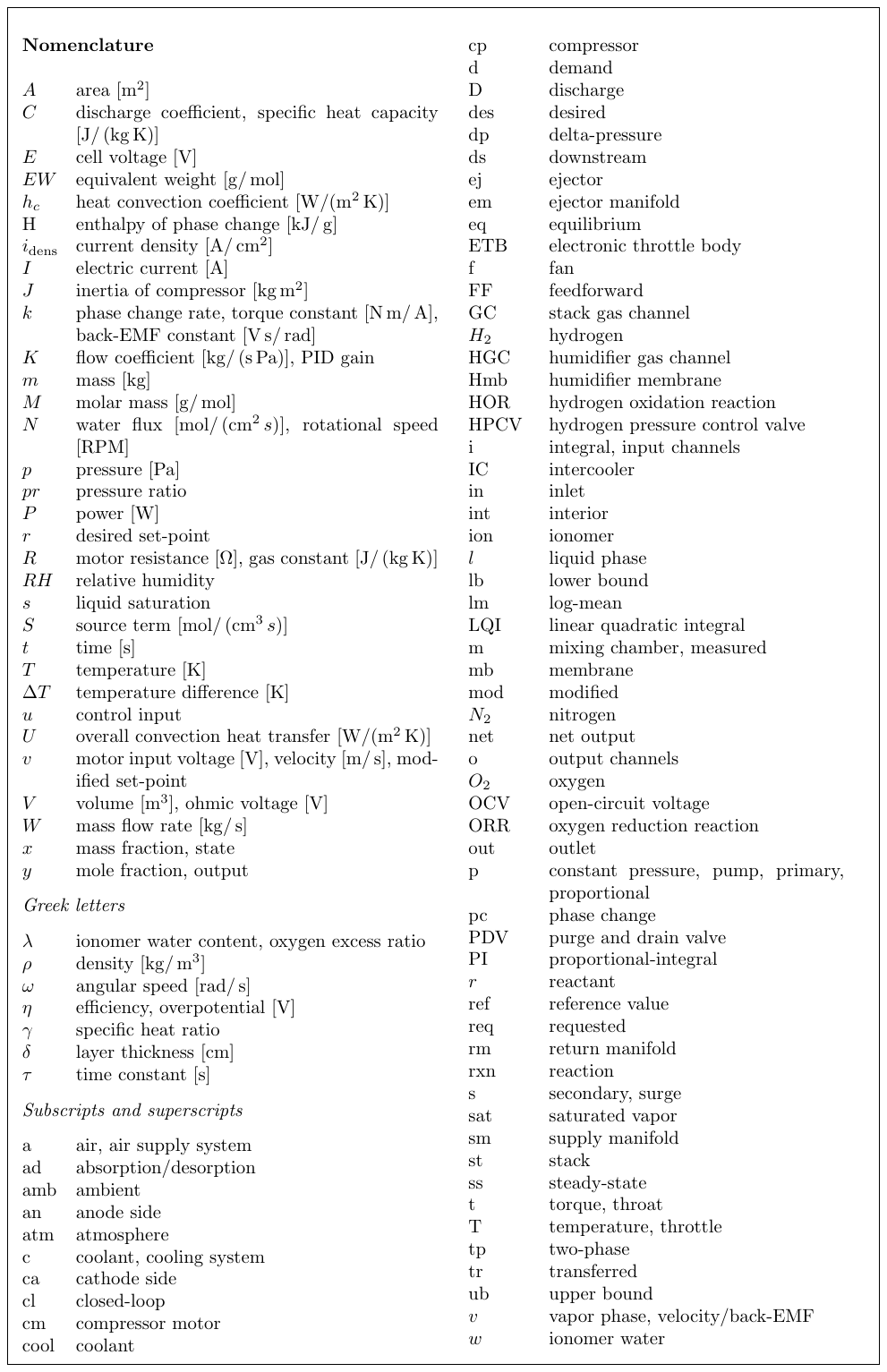}
\caption*{}
\label{fig:kjk}
\end{figure}
The increasing demand for clean and efficient transportation has driven significant interest in alternative propulsion technologies that can reduce greenhouse gas emissions and dependence on fossil fuels. Among these, proton exchange membrane fuel cells (PEMFCs) have emerged as a promising solution for automotive applications due to their high energy efficiency, low operating temperature, rapid startup capability, and zero tailpipe emissions \cite{qasem2024recent,abedin2025proton}. However, the widespread adoption of PEMFC-powered vehicles remains limited by several technical and economic challenges, including high system cost and limited durability \cite{whiston2019expert,chu2022experimental}. Although several material-based solutions, such as reversal-tolerant anode catalysts \cite{hong2016impact,mandal2018understanding}, membrane additives, and structural reinforcements \cite{kienitz2011ultra,stewart2014ceria}, have been used to improve the durability of the fuel cell stack, they often increase the overall cost of the fuel cell system (FCS). As a result, active control strategies have gained attention as a complementary means to meet both cost and durability targets in automotive PEMFC systems.

An automotive FCS typically consists of several interconnected subsystems, including the fuel cell stack where electrochemical reactions take place, the air supply system, the fuel supply system, the cooling system, and the power conditioning system (see Fig.~\ref{fig:FCS}). In the fuel cell control literature, the primary focus has been on regulating the operating conditions of the FCS to maximize net power output during power delivery. For the air supply subsystem, the control-oriented model presented in \cite{pukrushpan2004control3} is widely used as a basis for controller design. The purpose of the air supply controller is to ensure high net power delivery of the FCS by minimizing the compressor’s parasitic power consumption while supplying sufficient oxygen to the cathode. This is typically achieved by tracking optimal set-points for the compressor flow rate and cathode supply manifold pressure using various control strategies~\cite{pukrushpan2004control3,bacher2022efficiency,pilloni2015observer}.
In parallel with the air supply system, the efficiency of the FCS also depends on the thermal condition of the stack. The objective of the cooling system controller is to regulate the stack temperature while minimizing the power consumption of the radiator fan and coolant pump.  In \cite{chen2020robust}, the optimal coolant outlet temperature was determined as a function of operating conditions to maximize the FCS efficiency, and a cascade Internal Model Control (IMC) strategy was proposed to solve the tracking problem. In \cite{yu2008thermal}, the optimal coolant inlet and outlet temperatures were determined based on a trade-off between the effect of temperature distribution on fuel cell performance and the parasitic losses associated with the coolant pump and radiator fan \cite{ayubirad2024model}. A feedback control strategy was also developed to track the resulting optimal coolant inlet and outlet temperatures.

FCSs can experience durability issues, for which various constraint-aware control frameworks have been explored in the literature to address them. To protect the FCS against oxygen starvation in the cathode, the model predictive control (MPC) strategy was implemented in \cite{gruber2012nonlinear,gruber2009design, arce2009real, 11151331}. MPC has also been applied in experimental setups to simultaneously
reduce hydrogen consumption, prevent oxygen starvation, and track the desired load profile \cite{ziogou2013empowering,ziogou2013line}. As a computationally efficient alternative, the Reference Governor (RG) strategy was employed in \cite{sun2005load,vahidi2006constraint, 11049058} to enforce constraints associated with compressor surge, choke, and oxygen starvation by slowing down the transitions in current demand as little as possible. A novel multi-reference cascaded cross-section RG strategy was developed in \cite{bacher2023hierarchical}, which improved oxygen starvation constraint satisfaction by modifying, if necessary, both the optimal set-points of the air supply system and the current demand. However, these constraint-aware strategies are limited in their ability to predict the behavior of the fuel cell stack, as they rely on simplified models and disregard the intricate physical phenomena occurring within the cell. As a result, they fall short in addressing critical longevity challenges such as membrane flooding and dry-out \cite{singh20173d}. 

Several works have focused on closed-loop water management approaches that rely on real-time measurement of water content within the fuel cell stack. Among these, impedance-based techniques, particularly electrochemical impedance spectroscopy (EIS), have been widely used due to their sensitivity to hydration-related phenomena such as membrane flooding and dry-out \cite{najafi2020rapid,xu2022closed,kurz2008impedance}.
Although these strategies are effective in responding to hydration faults, they rely on measured deviations from nominal behavior, limiting their ability to predict and prevent degradation under dynamic operating conditions.

To support the development of predictive, constraint-aware control strategies for addressing membrane durability concerns, researchers have developed cell-level models that provide detailed information about water distribution throughout the membrane while striking a balance between fidelity and computational cost \cite{goshtasbi2019modeling}. In \cite{goshtasbi2020degradation}, a one-dimensional (1D) model of the cell was used to simulate water transport through the membrane thickness,  while the rest of the plant was modeled using component models adopted from existing literature\cite{pukrushpan2004control3}. This integrated model served as the basis for a linear time-varying model predictive controller (LTV-MPC), which enforced constraints on compressor and fuel cell variables, including temperature and membrane hydration. Specifically, the controller influenced membrane hydration by adjusting anode and cathode inlet relative humidity, coolant inlet temperature, and coolant flow rate through humidifier and coolant control. However, in state-of-the-art automotive FCS, the incoming air is passively humidified, and the inlet relative humidity of the anode and cathode is not directly controllable for membrane hydration management.
Also, due to the slow response of the thermal system and potential actuator saturation under hot ambient conditions, coolant control may not always be available to the extent that is desirable for hydration management. Therefore, our earlier work \cite{ayubirad2023simultaneous} treated the coolant inlet temperature and flow rate as fixed inputs and instead focused on managing membrane hydration through the air supply and current control in a 1D PEMFC model during both steady-state and transient conditions. In that work, the model in \cite{goshtasbi2020degradation} was extended with gas channel and coolant thermal dynamics, while the balance-of-plant (BoP) components were adopted from \cite{pukrushpan2004control3} to develop a parallel cascaded RG that maintained membrane hydration within safe bounds by modifying, when necessary, the current demand and the air supply set-point.

While these works lay a good foundation for designing controllers that address durability concerns at the stack-level, the current gaps in this field can be summarized as follows:
\begin{itemize}
    \item The existing models in fuel cell control literature fall short of capturing the distribution of stack variables along the flow channels, which limits their ability to represent local hydration and thermal phenomena.
    \item Prior works often use simplified representations of BoP components; for example, the humidifier is modeled as an ideal water injector without internal dynamics, which limits the evaluation of realistic water management strategies.
    \item The cooling system (radiator, fan, coolant pump) is typically excluded from the model structure, further reducing the utility of prior models for combined thermal and hydration control studies.
    \item The existing models have little to no validation at the component- and system-level and show limited fidelity in quantifying the impact of control inputs on heat and water distribution in the stack.
    \item The existing hydration management studies focus on spatially averaged or through-plane water dynamics, while control of hydration gradients along the flow channels remains unexplored. This is a key gap, as dry-out and flooding are localized phenomena, and membrane failure in one region can lead to stack-level failure.
\end{itemize}

Following the above discussion and to address these modeling and control gaps in the fuel cell literature, we first present a comprehensive dynamic model of an FCS that captures the distributions of hydration and thermal states along the flow channels. Specifically, the model integrates a pseudo-two-dimensional (P2D) representation of the stack with detailed dynamics of all BoP components used in state-of-the-art automotive FCS architectures. Building on this model, we then propose a constraint-aware control strategy to manage localized hydration of the stack under varying operating conditions. The main objectives of these controllers are to maximize net power output while protecting the FCS from operational constraint violations that could compromise performance or durability of the system. These objectives are addressed through a two-tier control design methodology. First, set-point maps for the air supply and cooling systems are generated to maximize net power output while satisfying all relevant constraints at steady-state. These include oxygen starvation, membrane hydration, and compressor surge and choke limits. As for the transient behavior, in this paper we focus only on membrane dry-out and compressor surge constraints to address the potential violations that can occur during load reductions. The constraint satisfaction during transients is achieved by modifying, when necessary, the optimal set-points of the air supply control system.
Although membrane hydration is more sensitive to changes in stack temperature~\cite{xu2022closed}, the cooling system's contribution to hydration management during transients is limited due to its slow response. Therefore, this paper explores the use of air supply system, characterized by its faster dynamic response, as a means of mitigating constraint violations during transient operation. To this end, a high-fidelity closed-loop FCS model is used to design a predictive control strategy within the Command Governor (CG) framework \cite{garone2017reference} for managing membrane hydration at the anode inlet, the driest region of the stack in a counter-flow configuration\footnote{In a counter-flow configuration, the hot and cold fluids flow in opposite directions to maximize the temperature gradient and heat transfer efficiency.}.

In summary, the novelties of this study can be listed as follows:
\begin{itemize}
\item Presentation of a comprehensive, integrated model of an automotive FCS that captures the nonlinear dynamics and coupling between the fuel cell stack and all the major BoP components, addressing the limitations of prior works that model only a subset of subsystems in detail while simplifying others.
\item Development of a fast-tracking and easily tunable linear quadratic integral (LQI) controller for the air supply system, with a focus on mitigating aggressive control inputs while maintaining the desired tracking performance.
\item Nonlinearity analysis of hydration dynamics at the driest part of the stack, followed by the derivation of a single reduced-order linearized model with sufficient fidelity to enable implementation of computationally efficient dry-out constraint enforcement.
\item Development of a constraint-aware control strategy within the CG framework for localized hydration management by adjusting the air supply set-points during power transitions when necessary.
\item Implementation of a CG which is modified from the existing literature by shortening the prediction horizon and softening the constraints to enable real-time feasibility, followed by evaluation on a proprietary high-fidelity model provided by Ford Motor Company using real-world drive cycles.
\end{itemize}
To the best of our knowledge, this study is the first to develop and validate a system-level constraint-aware control strategy that addresses stack-level durability concerns using a high-fidelity model in which the stack and all the major BoP subsystems are comprehensively validated against both component- and system-level experimental data, as presented in \cite{goshtasbi2021model}. 

\begin{figure}[t!]
\centering
\includegraphics[width=0.8\textwidth]{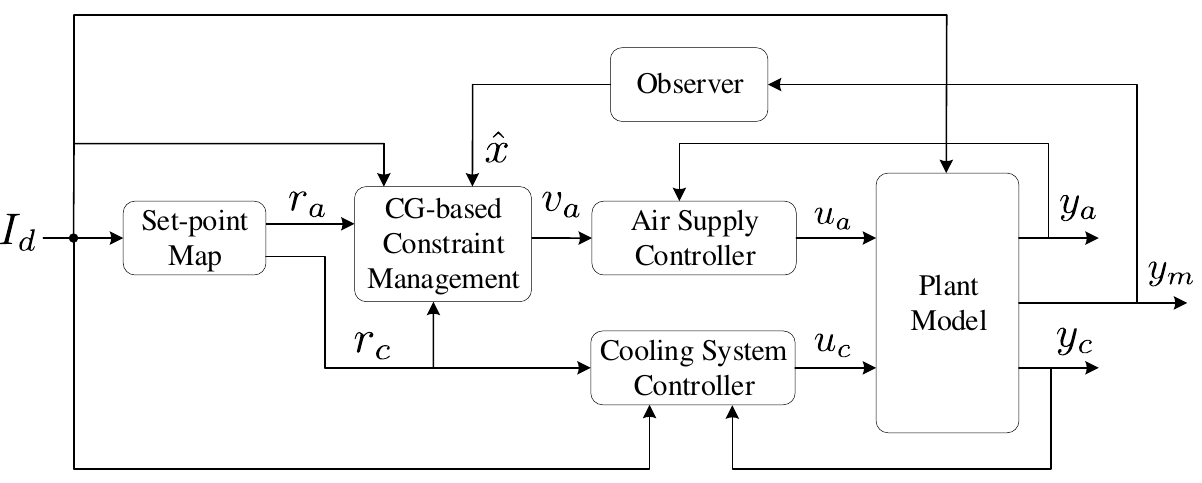}
\caption{Block diagram of the proposed constraint management strategy for localized water management. $I_\text{d}$ is the current demand input. $r_\text{a}$ and $r_\text{c}$ denote the desired set-points for the air and coolant supply subsystems, while $v_\text{a}$ is the adjusted set-point provided to the air supply controller. $u_\text{a}$ and $u_\text{c}$ are the control actions applied to the subsystems, and $y_\text{a}$ and $y_\text{c}$ denote the corresponding subsystem outputs. $\hat{x}$ denotes the internal states associated with membrane hydration, estimated by an observer from $y_\text{m}$.}
\label{fig:total_scheme}
\end{figure}

The proposed constraint-aware control strategy is summarized in the block diagram shown in Fig.~\ref{fig:total_scheme}. The definitions of the signals in Fig.~\ref{fig:total_scheme} are as follows:  
The current demand $I_{\text{d}}$ is computed to meet the requested power, with the power request mapped to $I_{\text{d}}$ by a high-level controller not shown in the figure. The outputs of the air supply system, $y_\text{a}$, are the compressor flow rate ($W_{\text{cp}}$) and the cathode supply manifold pressure ($p_{\text{sm}}^{\text{ca}}$). The outputs of the coolant supply system, $y_\text{c}$, are the coolant inlet ($T_{\text{c,in}}$) and outlet ($T_{\text{c,out}}$) temperatures. The desired set-points $r_\text{a} = [W_{\text{cp}}^{\text{des}},\, p_{\text{sm,ca}}^{\text{des}}]^\top$ and \mbox{$r_\text{c} = [T_{\text{c,in}}^{\text{des}},\, T_{\text{c,out}}^{\text{des}}]^\top$} are determined based on $I_{\text{d}}$ using a set-point map. In the CG-based constraint management block, the CG modifies the pressure set-point during transients to satisfy the hydration constraint while a heuristic approach adjusts the flow rate set-point to prevent surge, yielding the applied input $v_a = [\,W_{\mathrm{cp}}^{\mathrm{mod}},\; p_{\mathrm{sm,ca}}^{\mathrm{mod}}\,]^\top$. The control inputs, grouped into the vectors $u_\text{a}$ and $u_\text{c}$ for the air supply and cooling systems, respectively, include the compressor motor speed reference $N_{\text{cp,ref}}$ and throttle valve position $u_{\text{ETB}}$ in $u_\text{a}$, and the coolant pump voltage $v_\text{p}$ and radiator fan voltage $v_\text{f}$ in $u_\text{c}$. The measured outputs $y_\text{m}$ are sensor signals used by an observer to estimate the internal states $\hat{x}$ associated with membrane hydration.

The rest of the paper is organized as follows. Section ~\ref{sec:Model Development} describes the comprehensive model of the FCS. Section~\ref{sec:Air-supply Subsystem Control Development} presents the controller design for the FCS. In Section~\ref{sec:Hydration_constraint_management}, hydration constraint enforcement strategy is developed and evaluated using power trajectories from realistic drive cycles. Finally, Section~\ref{sec:Conclusion} concludes the paper with key remarks.

\section{Plant Model}
\label{sec:Model Development}
This section presents a comprehensive dynamic model of an FCS, referred to herein as the plant model. While, existing studies provide detailed models of the fuel cell stack \cite{promislow2008two, ziegler2005two, wu2010steady} and individual BoPs \cite{he2011analysis,pukrushpan2004control}, integrated models of the FCS in the literature often focus on a subset of system components while treating the remaining ones in a simplified manner \cite{kim2020parametric, kim2016parametric, yang2019comprehensive}. A comprehensive integrated model that accounts  for the nonlinear and complex interactions among these components is necessary for quantifying the impact of practical control actions on heat and water distribution in the stack, thereby enabling the design and validation of dynamic water balance control strategies. The contribution of this section is not in developing new models, as the stack and the BoP component models are adopted from existing literature. However, to the best of the authors’ knowledge, the study of interactions between the PEMFC stack and the various subsystems that incorporate both thermal and two-phase dynamics remains limited, a gap that this section seeks to fill.

\begin{figure}[htbp]
\centering
\includegraphics[width=1.0\textwidth]{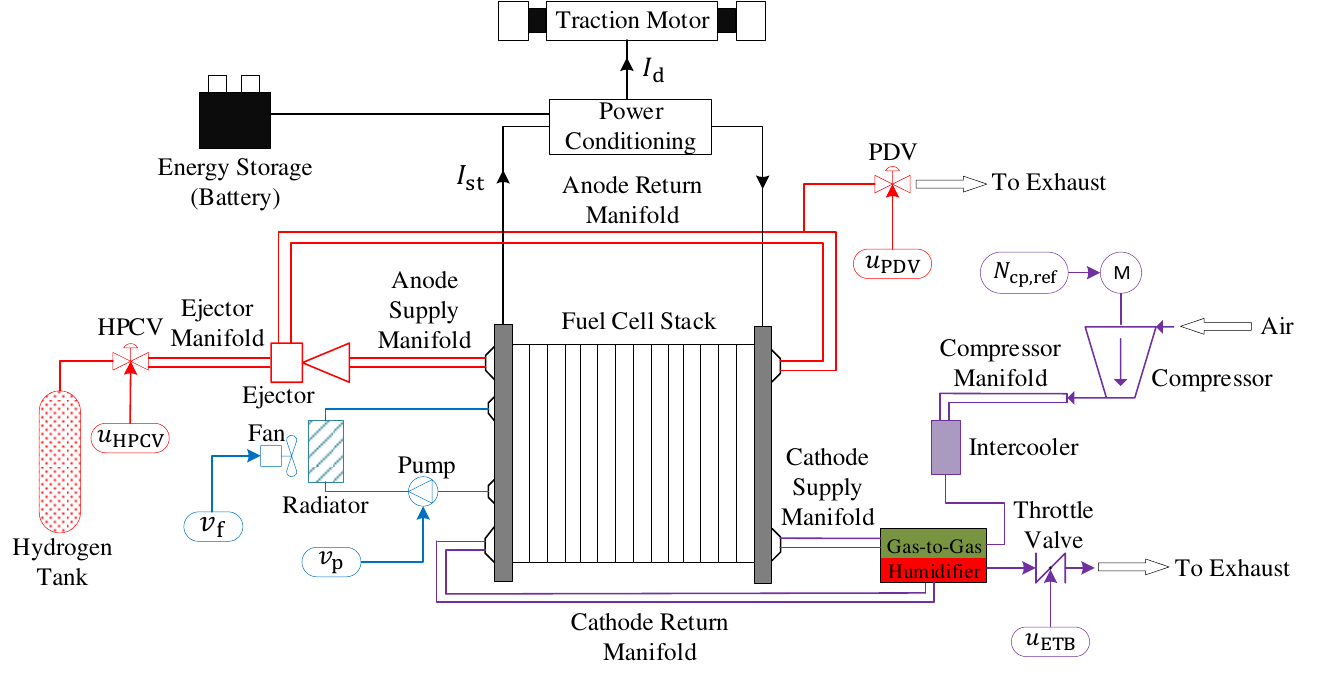}
\caption{Schematic diagram of the fuel cell system. All volumes and manifolds with associated dynamic states are identified by double lines. Single lines indicate direct connections between components without intermediate volumes. The red section represents the anode-side BoP components, the purple section corresponds to the cathode-side BoP components, and the blue section illustrates the cooling system.}
\label{fig:FCS}
\end{figure}

Based on the architecture illustrated in Fig.~\ref{fig:FCS}, the BoP is organized into three major subsystems for modeling and control purposes: the air supply system, the fuel delivery system, and the coolant supply system, each with its own control objective. In the following sections, a non-isothermal model of all BoP components incorporating two-phase dynamics is presented, using the same modeling approach used in our previous publication on the PEMFC stack \cite{ayubirad2023simultaneous}.  While most equations and submodels are adopted from existing literature and cited accordingly, some of the BoP models are extended to account for additional heat and mass transport processes in order to improve the model’s physical accuracy. For the stack and the coolant supply system, the reader is referred to previous publications \cite{ayubirad2023simultaneous,goshtasbi2019modeling}, where the modeling details and their interactions are thoroughly presented.

For model validation, a ``high-fidelity'' model is obtained by tuning the plant parameters to closely match experimental measurements. The high-fidelity model used in this study is proprietary to Ford Motor Company, with the stack and all sub-models rigorously validated at both the component- and system-level, under steady-state and transient operating conditions~\cite{goshtasbi2021model}.

\subsection{Air supply System }
\label{sec:Air-supply System Model}
The key functions of the air supply system are to supply sufficient oxygen to the stack and to support proper membrane hydration. As shown in Fig.~\ref{fig:FCS}, ambient air enters the compressor, where it is pressurized, resulting in an increase in temperature. This high-temperature, pressurized air then passes through an intercooler to reduce its temperature before entering the dry side of the humidifier. The cooled, pressurized air is humidified by moisture from the stack exhaust, which flows through the wet side of the humidifier. This configuration enables passive transfer of humidity from the exhaust to the incoming air. A throttle valve located downstream of the humidifier, together with the compressor, is actively controlled to regulate both mass flow rate and operating pressure of the cathode loop.

\subsubsection{Air compressor}
\label{sec:Air Compressor}
The air compressor model consists of two parts that describe the static characteristics and the dynamic behavior of the system. To compute the compressor efficiency and its outlet flow rate, static analytical expressions are used. For the compressor flow rate, we adopt the curve-fitting approach proposed by Jensen and Kristensen \cite{pukrushpan2004control3}, in which the output mass flow rate $W_{\mathrm{cp}}$ is formulated as a function of the pressure ratio across the compressor and the compressor motor speed as:
\begin{equation}
    W_{\mathrm{cp}} = f_{\mathrm{cp}}(pr, \omega_{\mathrm{cm}}, p_{\mathrm{atm}}, T_{\mathrm{atm}}),
\end{equation}
where $pr = p_{\mathrm{cp,ds}} / p_{\mathrm{atm}}$ is the pressure ratio across the compressor, $p_{\mathrm{cp,ds}}$ is the downstream pressure of the compressor, and $\omega_{\mathrm{cm}}$ is the compressor motor speed. The analytical expression of $f_{\mathrm{cp}}$ used in this work follows the formulation provided in \cite{pukrushpan2004control3}, Eqs.~(2.1)–(2.7). Similarly, the compressor efficiency $\eta_{\mathrm{cp}}$ is expressed using a polynomial surface, a form widely adopted in both industry and research \cite{marchante2023critical}, and fitted to a large dataset of operating points. The efficiency is described as a function of pressure ratio and compressor flow rate:
\begin{equation}
    \eta_{\mathrm{cp}} = h_{\mathrm{cp}}(pr, W_{\mathrm{cp}}).
\end{equation}
These formulations enable smooth and differentiable mapping of flow and efficiency, making the model suitable for control applications involving linearization. 

In addition to the static relationships, the model includes three dynamic states: compressor speed $\omega_{\mathrm{cm}}$, outlet air temperature $T_{\mathrm{cp}}$, and compressor downstream pressure $p_{\mathrm{cp,ds}}$, whose dynamics are described by the following equations:
\begin{equation}
\label{eqn:compressor_speed}
    \frac{d\omega_{\mathrm{cm}}}{dt} = \frac{k_t \eta_{\mathrm{cm}}}{J_{\mathrm{cp}} R_{\mathrm{cm}}} 
    (v_{\mathrm{cm}} - k_v \omega_{\mathrm{cm}}) 
    - \frac{C_{p,a}}{J_{\mathrm{cp}} \eta_{\mathrm{cp}}} 
    \frac{T_{\mathrm{atm}} W_{\mathrm{cp}}}{\omega_{\mathrm{cm}}} 
    \left[ \left( \frac{p_{\mathrm{cp,ds}}}{p_{\mathrm{atm}}} \right)^{\frac{\gamma_a - 1}{\gamma_a}} - 1 \right],
\end{equation}
\begin{equation}
\label{eqn:compressor_temp}
    \frac{dT_{\mathrm{cp}}}{dt} = \frac{T_{\mathrm{cp,ss}} - T_{\mathrm{cp}}}{\tau_{T,\mathrm{cp}}},
\end{equation}
\begin{equation}
\label{eqn:compressor_pressure}
    \frac{dp_{\mathrm{cp,ds}}}{dt} = \frac{R_a T_{\mathrm{cp}}}{V_{\mathrm{cp,ds}}} 
    (W_{\mathrm{cp}} - W_{\mathrm{cp,ds,out}}),
\end{equation}
where $W_{\mathrm{cp,ds,out}}$ and $T_{\mathrm{cp,ss}}$  represent the outlet flow from the compressor downstream volume and the compressor outlet temperature at steady-state, respectively, and are calculated as follows: 
\begin{equation}
\label{eqn:compressor_manifold_flow}
    W_{\mathrm{cp,ds,out}} = K_{\mathrm{dp,cp,ds}} (p_{\mathrm{cp,ds}} - p_{\mathrm{a,IC}}),
\end{equation}
\begin{equation}
    T_{\mathrm{cp,ss}} = T_{\mathrm{atm}} + \frac{T_{\mathrm{atm}}}{\eta_{\mathrm{cp}}} 
    \left[ \left( \frac{p_{\mathrm{cp,ds}}}{p_{\mathrm{atm}}} \right)^{\frac{\gamma_a - 1}{\gamma_a}} - 1 \right].
\end{equation}
Here, Eq.~\eqref{eqn:compressor_speed}, corresponding to Eqs.~(2.9--2.11) in~\cite{pukrushpan2004control3}, describes the compressor motor’s angular velocity dynamics based on the balance between applied motor torque and the torque required for gas compression. Eq.~\eqref{eqn:compressor_temp} models the non-isothermal behavior of the compressor outlet flow using first-order lag dynamics, while Eq.~\eqref{eqn:compressor_pressure} is derived from the dynamic mass conservation of the compressor downstream volume. Finally, Eq.~\eqref{eqn:compressor_manifold_flow} models the outlet flow rate from the compressor downstream volume using a linear pressure-drop relationship, which is a valid approximation when the pressure difference is small~\cite{pukrushpan2004control3}. In Eq.~\eqref{eqn:compressor_manifold_flow}, $p_{\mathrm{a,IC}}$ denotes the air pressure in the intercooler, whose dynamics are given by Eq.~\eqref{eqn:Inter_pa} in Section~\ref{sec:Intercooler}.

In this work, similar to \cite{liu2016modeling,chen2018control}, reference speed is considered as the input control of the air compressor model. This modeling choice reflects the use of a cascaded control architecture, where the inner loop controls the compressor motor voltage $v_{\mathrm{cm}}$ based on feedback from the actual compressor speed. The outer loop generates the speed reference $N_{\mathrm{cp,ref}}$ to control the air flow rate and pressure in the cathode loop.

\subsubsection{Intercooler}
\label{sec:Intercooler}
The intercooler is placed after the compressor to reduce the temperature of the compressed air before it enters downstream components. Acting as a heat exchanger, it transfers energy from the hotter compressed air to the cooler liquid coolant without phase change. The thermal interaction, with the intercooler operating in a counter-flow configuration, is modeled under steady-state energy balance assumption, where the heat rejected by the air is equal to the heat absorbed by the coolant:
\begin{equation}
\label{eq:intercooler_ss}
    W_{\mathrm{a,IC,in}} C_{p,a} (T_{\mathrm{a,IC,in}} - T_{\mathrm{a,IC,ss}}) = 
    W_{\mathrm{cool,IC}} C_{p,\mathrm{cool}} (T_{\mathrm{c,IC,ss}} - T_{\mathrm{c,in}}) = 
    U_{\mathrm{IC}} A_{\mathrm{IC}} \Delta T_{\mathrm{IC,lm}}.
\end{equation}
Here, $W_{\mathrm{cool,IC}}$ and $T_{\mathrm{c,in}}$ are the measured inlet flow rate and temperature of the coolant, while the air-side inlet flow rate and temperature are provided by the upstream compressor model. The log-mean temperature difference for the counter-flow configuration is also calculated as:
\begin{equation}
    \Delta T_{\mathrm{IC,lm}} = 
    \frac{(T_{\mathrm{a,IC,in}} - T_{\mathrm{c,IC,ss}}) - (T_{\mathrm{a,IC,ss}} - T_{\mathrm{c,in}})}%
         {\ln \left( \frac{T_{\mathrm{a,IC,in}} - T_{\mathrm{c,IC,ss}}}%
         {T_{\mathrm{a,IC,ss}} - T_{\mathrm{c,in}}} \right)}.
\end{equation}
In Eq.~\eqref{eq:intercooler_ss}, the outlet temperatures $T_{\mathrm{a,IC,ss}}$ and $T_{\mathrm{c,IC,ss}}$ are unknown and are therefore computed either numerically via Newton iteration or analytically using the effectiveness–NTU method \cite{incropera1996fundamentals}, depending on the implementation. Once the steady-state temperatures are determined, the non-isothermal behavior of the air and coolant is modeled using first-order lag dynamics:
\begin{equation}
    \frac{dT_{\mathrm{a,IC}}}{dt} = \frac{T_{\mathrm{a,IC,ss}} - T_{\mathrm{a,IC}}}{\tau_{T,\mathrm{IC}}},
\end{equation}
\begin{equation}
    \frac{dT_{\mathrm{c,IC}}}{dt} = \frac{T_{\mathrm{c,IC,ss}} - T_{\mathrm{c,IC}}}{\tau_{T,\mathrm{IC}}}.
\end{equation}
In parallel, air mass conservation is used to describe the evolution of pressure in the intercooler volume:
\begin{equation}
\label{eqn:Inter_pa}
    \frac{dp_{\mathrm{a,IC}}}{dt} = \frac{R_a T_{\mathrm{a,IC,lm}}}{V_{\mathrm{IC}}} 
    \left( W_{\mathrm{a,IC,in}} - W_{\mathrm{a,IC,out}} \right),
\end{equation}
where \( T_{\mathrm{a,IC,lm}} \) is the log-mean temperature of the air between \( T_{\mathrm{a,IC,in}} \) and \( T_{\mathrm{a,IC,ss}} \). The air outlet flow rate is described by a linear pressure-drop relationship:
\begin{equation}
    W_{\mathrm{a,IC,out}} = K_{\mathrm{dp,IC}} (p_{\mathrm{a,IC}} - p_{\mathrm{HGC}}^{\mathrm{dry}}),
\end{equation}
where $p_{\mathrm{HGC}}^{\mathrm{dry}}$ is pressure on the dry side of the humidifier, whose dynamics are described in Section~\ref{sec:Humidifier}. 

\subsubsection{Shell-and-tube type gas-to-gas membrane humidifier}
\label{sec:Humidifier}
\begin{figure}[t!]
\centering
\includegraphics[width=0.8\textwidth]{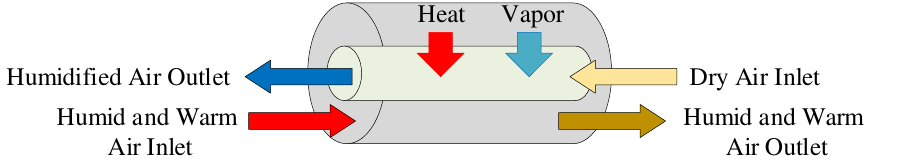}
\caption{Schematic of a shell-and-tube type gas-to-gas membrane humidifier.}
\label{fig:Shell_Tube}
\end{figure}

A shell-and-tube type gas-to-gas membrane humidifier is commonly used for external humidification of fuel cell reactant gases in mobile applications because of its compact design, lack of moving parts, and the fact that it requires no additional power supply \cite{park2008dynamic}. In this configuration, dry air from the intercooler flows through the membrane tubes (dry chennel), while humid and warm exhaust air from the fuel cell stack circulates around the tubes within the shell (wet channel), as shown in Fig.~\ref{fig:Shell_Tube}. The passive transfer of heat and water vapor between the fluids across the membrane humidifies and preheats the dry inlet air without the need for active components.

As a key component in water management, an accurate dynamic model of the humidifier is essential for predicting membrane hydration and control system development. Perhaps the most comprehensive shell-and-tube type gas-to-gas membrane humidifier model was proposed in \cite{yang2019comprehensive}, as it incorporates membrane water transport along with the  non-isothermal and two-phase channel dynamics.

The dynamics of the gas streams in the dry and wet channels of the humidifier are obtained by solving the dynamic conservation of mass, as follows::
\begin{equation}
\label{eqn:hum_vapor_p}
\frac{d p_v^i}{dt} = \frac{R_v T_{\mathrm{HGC}}^i}{V_{\mathrm{HGC}}^i (1 - s_{\mathrm{HGC}}^i)} 
\left( {W}_{v,\mathrm{in}}^i - {W}_{v,\mathrm{out}}^i + {W}_{v,\mathrm{tr}}^i - {W}_{\mathrm{pc}}^i \right) 
+ \frac{p_v^i}{T_{\mathrm{HGC}}^i} \frac{d T_{\mathrm{HGC}}^i}{dt},
\end{equation}
\begin{equation}
\label{eqn:hum_air_p}
\frac{d p_a^i}{dt} = \frac{R_a T_{\mathrm{HGC}}^i}{V_{\mathrm{HGC}}^i (1 - s_{\mathrm{HGC}}^i)} 
\left( W_{a,\mathrm{in}}^i - W_{a,\mathrm{out}}^i \right) 
+ \frac{p_a^i}{T_{\mathrm{HGC}}^i} \frac{d T_{\mathrm{HGC}}^i}{dt},
\end{equation}
where the superscript \( i \in \{\mathrm{dry}, \mathrm{wet}\} \) indicates that the equation is applicable to both the dry and wet channels of the humidifier. Here, Eq.~\eqref{eqn:hum_vapor_p} is similar to Eq.~(30) in Ref.~\cite{yang2019comprehensive}, while Eq.~\eqref{eqn:hum_air_p} is not considered in that work, as the model assumes that the mass flow rate of dry air remains the same between the inlet and outlet of both the dry and wet channels of the humidifier. The liquid water saturation \( s_{\mathrm{HGC}} \) in Eqs.~\eqref{eqn:hum_vapor_p}--\eqref{eqn:hum_air_p} is defined as the ratio of liquid water volume to the total pore volume available for gas flow in the channel:
\begin{equation}
s_{\mathrm{HGC}}^i = \frac{m_{l,\text{HGC}}^i}{\rho_l V_{\mathrm{HGC}}^i},
\end{equation}
where \( m_{l,\text{HGC}} \) is the mass of accumulated liquid water in the channels. The mass flow rates of the inlet water vapor and dry air streams into the dry channel of the humidifier are derived using the upstream intercooler outlet flow rate, as follows:
\begin{equation}
W_{v,\mathrm{in}}^{\mathrm{dry}} = x_{v,\mathrm{atm}} \, W_{a,\mathrm{IC,out}},
\end{equation}
\begin{equation}
W_{a,\mathrm{in}}^{\mathrm{dry}} = (1 - x_{v,\mathrm{atm}}) \, W_{a,\mathrm{IC,out}},
\end{equation}
while the wet-side inlet flow rates correspond to the outlet flow rates of the cathode return manifold described in Section~\ref{sec:ca_Manifolds}. Similarly, the flow rates of water vapor and dry air at the outlet of the humidifier channels are calculated as \( W_{v,\mathrm{out}}^i = (1 - x_a^i) \, W_{\mathrm{out}}^i \) and \( W_{a,\mathrm{out}}^i = x_a^i \, W_{\mathrm{out}}^i \),  where $W_{\text{out}}^{\text{dry}}$ is given in Eq.~\eqref{eqn:dry_outlet_flow}, and 
$W_{\text{out}}^{\text{wet}}$ is defined in Eq.~\eqref{eqn:wet_outlet_flow}.
\begin{equation}
\label{eqn:dry_outlet_flow}
W_{\mathrm{out}}^{\mathrm{dry}} = K^{\mathrm{dry}}_{\mathrm{dp,HGC}} \left( p^{\mathrm{dry}}_{\mathrm{HGC}} - p_{\mathrm{sm}}^{\mathrm{ca}} \right),
\end{equation}
\begin{footnotesize}
\begin{equation}
\label{eqn:wet_outlet_flow}
\begin{array}{l}
W_{\mathrm{out}}^{\mathrm{wet}} = \\[0.5ex]
\left\{
\begin{array}{ll}
\displaystyle
\frac{C_{\mathrm{D,ETB}} A_{\mathrm{T,ETB}}(u_{\mathrm{ETB}})\, p_{\mathrm{HGC}}^{\mathrm{wet}}}
     {\sqrt{R_{\mathrm{HGC}}^{\mathrm{wet}} T_{\mathrm{HGC}}^{\mathrm{wet}}}}
\left( \frac{p_{\mathrm{atm}}}{p_{\mathrm{HGC}}^{\mathrm{wet}}} \right)^{1/\gamma_{\mathrm{HGC}}^{\mathrm{wet}}}
\left\{ \frac{2\gamma_{\mathrm{HGC}}^{\mathrm{wet}}}{\gamma_{\mathrm{HGC}}^{\mathrm{wet}}-1}
        \left[ 1 -
        \left( \frac{p_{\mathrm{atm}}}{p_{\mathrm{HGC}}^{\mathrm{wet}}} \right)^{\frac{\gamma_{\mathrm{HGC}}^{\mathrm{wet}}-1}{\gamma_{\mathrm{HGC}}^{\mathrm{wet}}}} \right] \right\}^{1/2},
& \text{subsonic flow} \\[2ex]

\displaystyle
\frac{C_{\mathrm{D,ETB}} A_{\mathrm{T,ETB}}(u_{\mathrm{ETB}})\, p_{\mathrm{HGC}}^{\mathrm{wet}}}
     {\sqrt{R_{\mathrm{HGC}}^{\mathrm{wet}} T_{\mathrm{HGC}}^{\mathrm{wet}}}}
\,\sqrt{{\gamma_{\mathrm{HGC}}^{\mathrm{wet}}}}
\left( \frac{2}{\gamma_{\mathrm{HGC}}^{\mathrm{wet}} + 1} \right)^{\frac{\gamma_{\mathrm{HGC}}^{\mathrm{wet}} + 1}{2(\gamma_{\mathrm{HGC}}^{\mathrm{wet}} - 1)}},
& \text{choked flow}
\end{array}
\right.
\end{array}
\end{equation}
\end{footnotesize}
In Eq.~\eqref{eqn:wet_outlet_flow}, choked flow occurs when
\begin{equation*}
\frac{p_{\mathrm{atm}}}{p_{\mathrm{HGC}}^{\mathrm{wet}}} \leq 
\left( \frac{2}{\gamma_{\mathrm{HGC}}^{\mathrm{wet}} + 1} \right)^{\frac{\gamma_{\mathrm{HGC}}^{\mathrm{wet}}}{\gamma_{\mathrm{HGC}}^{\mathrm{wet}} - 1}}
\end{equation*}
and corresponds to subsonic flow otherwise.
Here, the outlet flow rate of the dry channel is modeled using a linear pressure-drop relationship, while the outlet flow from the wet channel of the humidifier is governed by a nozzle flow equation under the isentropic flow assumption~\cite{pukrushpan2004control3}. The pressures in the dry (\( p_{\mathrm{HGC}}^{\mathrm{dry}} \))and wet (\( p_{\mathrm{HGC}}^{\mathrm{wet}} \)) channels of the humidifier are the sum of the partial pressures of dry air and water vapor in the corresponding channel. In Eq.~\eqref{eqn:dry_outlet_flow}, \( p_{\mathrm{sm}}^{\mathrm{ca}} \) is the cathode supply manifold pressure, whose dynamics are described in Section~\ref{sec:ca_Manifolds}. In Eq.~\eqref{eqn:wet_outlet_flow}, the opening area of the throttle valve, connected to the wet channel of the humidifier, is modeled as a function of the control input \( u_{\mathrm{ETB}} \) \cite{kim2016parametric}, which represents the valve opening percentage in the range \([0,\,100]\).
The rate of liquid condensation or evaporation in Eq.~\eqref{eqn:hum_vapor_p} is calculated using Eq.~(13) from Ref.~\cite{ayubirad2023simultaneous}:
\begin{equation}
W_{\mathrm{pc}}^i = k_{\mathrm{pc}} V_{\mathrm{HGC}}^i \, \frac{p_{v,\mathrm{sat}}(T_{\mathrm{HGC}}^i)}{R_v T_{\mathrm{HGC}}^i} \left( RH_{\mathrm{HGC}}^i - 1 \right),
\end{equation}
where \( p_{v,\mathrm{sat}} \) is the saturation pressure, which depends on the temperature of the channels and is computed using Eq.~(2.10) of Ref.~\cite{goshtasbi2019modeling}, and \( RH_{\mathrm{HGC}} \) is the relative humidity, defined as:
\begin{equation}
RH_{\mathrm{HGC}}^i = \frac{p_v^i}{p_{v,\mathrm{sat}}(T_{\mathrm{HGC}}^i)}.
\end{equation}
In a membrane-based humidifier, water vapor from the wet gas stream is absorbed into the ionomer phase of the membrane, diffuses through it, and desorbs into the dry gas stream as vapor. Thus, the vapor transfer to and from the membrane in Eq.~\eqref{eqn:hum_vapor_p}, $W_{v,\mathrm{tr}}$, is determined by the water absorption/desorption source term \( S_{\mathrm{ad}} \) as shown in Table~A2 of Ref.~\cite{yang2019comprehensive}, and is calculated as follows:
\begin{equation}
W_{v,\mathrm{tr}}^i = - M_v A_{\mathrm{Hmb}} \, \delta_{\mathrm{Hmb}}^i \, S_{\mathrm{ad}}^i,
\end{equation}
where \( S_{\mathrm{ad}} \) is calculated using Eq.~(2.11) from Ref.~\cite{goshtasbi2019modeling}. The absorption/desorption source term is influenced by the water content in the membrane, as well as the relative humidity and temperature on both sides of the membrane. Accordingly, the thermal dynamic behavior of the membrane, as well as that of the dry and wet channels, is obtained by solving the dynamic conservation of energy equations and is calculated in a manner similar to Eqs.~(32)--(34) in Ref.~\cite{yang2019comprehensive}:
\begin{align}
\label{eqn:Hum_mem_T}
\frac{d T_{\mathrm{Hmb}}}{dt} = \,
& \frac{1}
{(\rho C_p)_{\mathrm{Hmb}} A_{\mathrm{Hmb}} \left( \delta_{\mathrm{Hmb}}^{\mathrm{int}} + \delta_{\mathrm{Hmb}}^{\mathrm{dry}} + \delta_{\mathrm{Hmb}}^{\mathrm{wet}} \right)}
\Bigg[
A_{\mathrm{Hmb}} \sum_i h_{c,\mathrm{Hmb}2i} \left( T_{\mathrm{HGC}}^i - T_{\mathrm{Hmb}} \right) \nonumber \\
&\quad - \sum_i \left\{
\sigma(W_{v,\mathrm{tr}}^i) W_{v,\mathrm{tr}}^i C_{p,v} T_{\mathrm{Hmb}} +
\left(1 - \sigma(W_{v,\mathrm{tr}}^i)\right) W_{v,\mathrm{tr}}^i C_{p,v} T_{\mathrm{HGC}}^i
\right\}
\Bigg],
\end{align}
\begin{align}
\label{eqn:Hum_GC_T}
\frac{d T_{\mathrm{HGC}}^i}{dt} = \,
& \frac{1}{\rho_{\mathrm{HGC}}^i V_{\mathrm{HGC}}^i C_{p,\mathrm{HGC}}^i (1 - s_{\mathrm{HGC}}^i)} \Big[
(W_{a,\mathrm{in}}^i C_{p,a} + W_{v,\mathrm{in}}^i C_{p,v}) T_{\mathrm{HGC,in}}^i \nonumber \\
& - (W_{a,\mathrm{out}}^i C_{p,a} + W_{v,\mathrm{out}}^i C_{p,v}) T_{\mathrm{HGC}}^i \nonumber  + \sigma(W_{v,\mathrm{tr}}^i) W_{v,\mathrm{tr}}^i C_{p,v} T_{\mathrm{Hmb}} \\
&+ (1 - \sigma(W_{v,\mathrm{tr}}^i)) W_{v,\mathrm{tr}}^i C_{p,v} T_{\mathrm{HGC}}^i 
+ W_{\mathrm{pc}}^i H_{\mathrm{pc}}(T_{\mathrm{HGC}}^i) \nonumber \\
& + h_{c,\mathrm{Hmb}2i} A_{\mathrm{Hmb}} (T_{\mathrm{Hmb}} - T_{\mathrm{HGC}}^i) 
+ h_{c,\mathrm{amb}2i} A_{\mathrm{amb}} (T_{\mathrm{amb}} - T_{\mathrm{HGC}}^i) \Big],
\end{align}
where
\begin{equation*}
\sigma(W_{v,\mathrm{tr}}^i) =
\begin{cases}
1, & \text{if } W_{v,\mathrm{tr}}^i > 0 \\
0, & \text{if } W_{v,\mathrm{tr}}^i \leq 0
\end{cases}
\end{equation*}
and \( H_{\mathrm{pc}} \) is the enthalpy of phase change, calculated using Eq.~(2.12) from Ref.~\cite{goshtasbi2019modeling}.
In Eqs.~\eqref{eqn:Hum_mem_T} and \eqref{eqn:Hum_GC_T}, the transient heat in the channels is influenced by the thermal energy of the inflowing and outflowing gases,  the change in internal energy due to convective heat transfer, and the heat associated with evaporation and condensation. Here, the formulations extend the model in Ref.~\cite{yang2019comprehensive} by also accounting for the heat transfer associated with the water vapor transport across the membrane.

The governing equations for the membrane water content, both within the membrane (\( \lambda_{\mathrm{Hmb}}^{\mathrm{int}} \)) and on its two sides (\( \lambda_{\mathrm{Hmb}}^i \)), are obtained by balancing water absorption/desorption and transport mechanisms, similar to the water content computation for the stack membrane in previous work (Eqs.~(2.49)--(2.51) in~\cite{goshtasbi2019modeling}), and are formulated in a manner similar to Eqs.~(28)--(29) in Ref.~\cite{yang2019comprehensive}:
\begin{equation}
\label{eqn:Hum_mem_dry}
\frac{d \lambda_{\mathrm{Hmb}}^{\mathrm{dry}}}{dt} = 
\frac{EW_{}}{\rho_{\mathrm{ion}}} 
\left( -\frac{N_w^{\mathrm{dry}2\mathrm{Hmb}}}{\delta_{\mathrm{Hmb}}^{\mathrm{dry}}} + S_{\mathrm{ad}}^{\mathrm{dry}} \right),
\end{equation}
\begin{equation}
\label{eqn:Hum_mem_int}
\frac{d \lambda_{\mathrm{Hmb}}^{\mathrm{int}}}{dt} = 
\frac{EW_{}}{\rho_{\mathrm{ion}} \delta_{\mathrm{Hmb}}^{\mathrm{int}}} 
\left( N_w^{\mathrm{dry}2\mathrm{Hmb}} - N_w^{\mathrm{Hmb}2\mathrm{wet}} \right),
\end{equation}
\begin{equation}
\label{eqn:Hum_mem_wet}
\frac{d \lambda_{\mathrm{Hmb}}^{\mathrm{wet}}}{dt} = 
\frac{EW_{}}{\rho_{\mathrm{ion}}} 
\left( \frac{N_w^{\mathrm{Hmb}2\mathrm{wet}}}{\delta_{\mathrm{Hmb}}^{\mathrm{wet}}} + S_{\mathrm{ad}}^{\mathrm{wet}} \right).
\end{equation}
This formulation also includes the explicit modeling of the membrane water content, which was not considered in Ref.~\cite{yang2019comprehensive}. In Eqs.~\eqref{eqn:Hum_mem_dry}--\eqref{eqn:Hum_mem_wet}, the interfacial water fluxes $N_{w}^{\mathrm{dry2Hmb}}$ and $N_{w}^{\mathrm{Hmb2wet}}$ are computed using Eq.~(2.7) in Ref.~\cite{goshtasbi2019modeling}, which accounts for diffusion, hydraulic permeation, and thermo-osmosis. This extends the approach in Ref.~\cite{yang2019comprehensive}, which considers only diffusion.

Finally, the dynamic equation for liquid water in the wet channel is obtained by applying the conservation of mass and is calculated similarly to Eq.~(31) in Ref.~\cite{yang2019comprehensive}, as follows:
\begin{equation}
\frac{d m_{l,\text{HGC}}^{\mathrm{wet}}}{dt} = W_{l,\mathrm{in}}^{\mathrm{wet}} - W_{l,\mathrm{out}}^{\mathrm{wet}} + W_{\mathrm{pc}}^{\mathrm{wet}},
\end{equation}
where the inlet liquid flow rate to the wet channel, \( W_{l,\mathrm{in}}^{\mathrm{wet}} \), is provided by the cathode return manifold model described in Section~\ref{sec:ca_Manifolds}, and the outlet liquid flow rate, \( W_{l,\mathrm{out}}^{\mathrm{wet}} \), is calculated using Eq.~(14) from Ref.~\cite{ayubirad2023simultaneous}:
\begin{equation}
W_{l,\mathrm{out}}^{\mathrm{wet}} = C_{\mathrm{tp}} \left( s_{\mathrm{HGC}}^{\mathrm{wet}} \right)^2 W_{\mathrm{out}}^{\mathrm{wet}}.
\end{equation}
As for the dry channel, liquid water is neglected since the water vapor concentration rarely reaches saturation. In the above equations, the gas mixture properties in the humidifier channels, including the heat capacity ratio \( \gamma_{\mathrm{HGC}} \), the specific gas constant \( R_{\mathrm{HGC}} \), the mixture density \( \rho_{\mathrm{HGC}} \), and the specific heat capacity \( C_{p,\mathrm{HGC}} \), are calculated using standard gas mixture relations~\cite{borgnakke2020fundamentals}.

\subsubsection{Cathode supply and return manifolds}
\label{sec:ca_Manifolds}
The dynamics of each manifold are described by differential equations involving five state variables: three partial pressures of the gas species, one liquid water mass, and one gas mixture temperature. The evolution of these states is governed by the conservation of mass and energy applied to each manifold:
\begin{equation}
\label{eqn:manifold_pr}
\frac{d p_r}{d t} = \frac{R_r T}{V(1 - s)} \left( W_{r,\text{in}} - W_{r,\text{out}} \right) + \frac{p_r}{T} \frac{dT}{d t},
\end{equation}
\begin{equation}
\label{eqn:manifold_pN}
\frac{d p_{{N}_2}}{d t} = \frac{R_{{N}_2} T}{V(1 - s)} \left( W_{{N}_2,\text{in}} - W_{{N}_2,\text{out}} \right) + \frac{p_{{N}_2}}{T} \frac{dT}{d t},
\end{equation}
\begin{equation}
\label{eqn:manifold_pv}
\frac{d p_v}{d t} = \frac{R_v T}{V(1 - s)} \left( W_{v,\text{in}} - W_{v,\text{out}} - W_{\text{pc}} \right) + \frac{p_v}{T} \frac{dT}{d t},
\end{equation}
\begin{equation}
\label{eqn:manifold_l}
\frac{d m_l}{d t} = W_{l,\text{in}} - W_{l,\text{out}} + W_{\text{pc}},
\end{equation}
\begin{equation}
\label{eqn:Manifold_T}
\frac{dT}{d t} = \frac{1}{\rho V C_p} \left\{ \sum_j C_{p,j} \left( W_{j,\text{in}} T_{\text{in}} - W_{j,\text{out}} T \right) + h_{\text{c,amb}} A_{\text{amb}} (T_{\text{amb}} - T) + W_{\text{pc}} H_{\text{pc}}(T) \right\}.
\end{equation}
Compared to the manifold models in \cite{pukrushpan2004control3}, the present model incorporates the non-isothermal and two-phase dynamics to provide a more detailed description of mass and heat transport. Here, the equations are formulated to apply across all manifolds. Accordingly, each variable adopts the appropriate indexing based on the manifold's type (supply or return) and location (anode or cathode). In these equations, the subscript $r \in \{{H}_2, {O}_2\}$ denotes the reactant gas species, and the subscript $j$ represents the gas components, where $j \in \{{O}_2, {N}_2, v\}$ on the cathode side and $j \in \{{H}_2, {N}_2, v\}$ on the anode side. 

On the cathode side, the inlet flow rates of the manifolds are determined by upstream components, including the dry outlet of the humidifier and the cathode outlet from the stack.
The outlet flow rates of the manifolds are modeled based on the pressure differences between each manifold and its corresponding downstream component:
\begin{equation}
W^{\mathrm{ca}}_{\mathrm{sm,out}} = K^{\mathrm{ca}}_{\mathrm{dp,sm}} \left( p_{\mathrm{sm}}^{\mathrm{ca}} - p_{\mathrm{GC}}^{\mathrm{ca}} \right),
\end{equation}
\begin{equation}
W^{\mathrm{ca}}_{\mathrm{rm,out}} = K^{\mathrm{ca}}_{\mathrm{dp,rm}} \left( p^{\mathrm{ca}}_{\mathrm{rm}} - p_{\mathrm{HGC}}^{\mathrm{wet}} \right),
\end{equation}
where \( p_{\mathrm{sm}}^{\mathrm{ca}} \) and \( p_{\mathrm{rm}}^{\mathrm{ca}} \) represent the total partial pressure of gas species in the cathode supply and return manifolds, respectively, and $p_{\mathrm{GC}}^{\mathrm{ca}}$ is the cathode gas channel pressure, whose gas species dynamics are described in Eqs.~(1)–(4) of our previous publication \cite{ayubirad2023simultaneous}. The species flow rates entering or exiting the manifolds in the above equations can be calculated from the total mass flow rate and species mass fractions as \( W_{j,k} = x_j W_k \), where \( k \in \{\text{in}, \text{out}\} \), and
\begin{equation}
\label{eqn:species _frac}
x_j = \frac{p_j M_j}{\sum_j p_j M_j}.
\end{equation}
where \( p_j \) and \( M_j \) denote the partial pressure and molar mass of species \( j \), respectively.
The calculation of gas mixture thermodynamic properties, phase change rates, and liquid flow rates within the manifold follows the same approach as in previous sections and is not repeated here.

\subsection{Fuel delivery system}
\label{subsec:FDS}
The fuel delivery system consists of a hydrogen pressure control valve (HPCV), an ejector, an ejector manifold, supply and return manifolds, and a purge valve, as shown in Fig.~\ref{fig:FCS}. The modeling of these components is detailed in the following sections.

\subsubsection{Hydrogen pressure control valve}
\label{sec:HPCV}
For modeling the HPCV, the valve is assumed to be connected to a high-pressure hydrogen storage tank. Due to the high upstream pressure, the HPCV flow is assumed to be choked and is modeled as isentropic nozzle flow, as given by Eq.~(2.17) in Ref.~\cite{pukrushpan2004control3}:
\begin{equation}
W_{\mathrm{HPCV}} = u_{\mathrm{HPCV}} \frac{C_{\mathrm{D},\mathrm{HPCV}} A_{\mathrm{T},\mathrm{HPCV}} p_{\mathrm{Tank}}}{\sqrt{R_{{H}_2} T_{\mathrm{Tank}}}} 
\gamma_{{H}_2}^{1/2} \left( \frac{2}{\gamma_{{H}_2} + 1} \right)^{\frac{\gamma_{{H}_2} + 1}{2(\gamma_{{H}_2} - 1)}},
\end{equation}
where \( u_{\mathrm{HPCV}} \) is the control input signal to the valve, and \( T_{\mathrm{Tank}} \) and \( p_{\mathrm{Tank}} \) are the measured temperature and pressure of the hydrogen storage tank, respectively.

\subsubsection{Ejector and ejector manifold}
\label{sec:ejector}
The ejector passively recirculates unreacted hydrogen from the anode exhaust back to the anode inlet. This recirculation enhances fuel utilization, increases overall system efficiency, and provides passive humidification of the incoming hydrogen. The lack of moving parts, zero power consumption, and passive operation make the ejector particularly suitable for mobile applications, similar to the shell-and-tube type gas-to-gas membrane humidifier described in Section~\ref{sec:Humidifier}. As shown in Fig.~\ref{fig:FCS}, the ejector consists of a primary and a secondary nozzle. High-pressure hydrogen from the HPCV flows through the primary nozzle, where its high velocity creates a low-pressure region that entrains the anode exhaust. The two streams then mix and continue toward the anode supply manifold. 

To model the ejector behavior, this study adopts a steady-state formulation developed in \cite{,he2011analysis,chen20131d}, which computes the primary (\(\text{p}\)) and secondary (\(\text{s}\)) flow rates under both subcritical and critical (choked) conditions, and determines the outlet temperature. For subcritical operation, where the pressure ratio is greater than the critical threshold, i.e.,
\begin{equation}
\frac{p_{\mathrm{ej,m}}}{p_{\mathrm{ej},i}} > \left( \frac{2}{\gamma_i + 1} \right)^{\gamma_i / (\gamma_i - 1)},
\end{equation}
the ejector flow rate is given by:
\begin{equation}
W_{\mathrm{ej},i} = A_{\mathrm{t},i} \sqrt{ \frac{2 \gamma_i}{\gamma_i - 1} \eta_i p_{\mathrm{ej},i} \rho_i \left[ \left( \frac{p_{\mathrm{ej,m}}}{p_{\mathrm{ej},i}} \right)^{2/\gamma_i} - \left( \frac{p_{\mathrm{ej,m}}}{p_{\mathrm{ej},i}} \right)^{(\gamma_i+1)/\gamma_i} \right] }, \quad i \in \{\text{p}, \text{s}\}
\end{equation}
where \( \gamma_i \) and \( \rho_i \) are the specific heat capacity ratio and density at the corresponding inlet, \( A_{\mathrm{t},i} \) is the nozzle throat area, \( \eta_i \) is the efficiency coefficient, and \( p_{\mathrm{ej,m}} \) is the mixing pressure. The term \( p_{\mathrm{ej},i} \) denotes the inlet pressure for each flow path, where \( p_{\mathrm{ej,p}} \) refers to the pressure of hydrogen in the ejector manifold, and \( p_{\mathrm{ej,s}} \) corresponds to the pressure in the anode return manifold. For critical (choked) flow, the mass flow rate simplifies to:
\begin{equation}
W_{\mathrm{ej},i} = A_{\mathrm{t},i} \sqrt{ \eta_i p_{\mathrm{ej},i} \rho_i \, \gamma_i \left( \frac{2}{\gamma_i + 1} \right)^{\frac{\gamma_i + 1}{\gamma_i - 1}} }.
\end{equation}
Once the primary and secondary flow rates are determined, the temperature of the mixed stream leaving the ejector is given by:
\begin{equation}
T_{\mathrm{ej},\mathrm{out},\mathrm{ss}} = \frac{W_{\mathrm{ej,p}} T_{\mathrm{ej},p} + W_{\mathrm{ej,s}} T_{\mathrm{ej,s}}}{W_{\mathrm{ej,p}} + W_{\mathrm{ej,s}}} - \frac{\gamma_\mathrm{m} - 1}{2 \gamma_\mathrm{m} R_\mathrm{m}} v_\mathrm{m}^2,
\end{equation}
with \( \gamma_\mathrm{m} \), \( R_\mathrm{m} \), and \( v_\mathrm{m} \) denoting the specific heat ratio, gas constant, and velocity after mixing, respectively. Here, \( T_{\mathrm{ej,p}} \) corresponds to the temperature of hydrogen in the ejector manifold, which is assumed equal to the tank temperature, and \( T_{\mathrm{ej,s}} \) is the temperature of the gas mixture in the anode return manifold. 

The dynamic behavior of the ejector is characterized by the hydrogen pressure in the adjacent manifold and the transient response of its outlet temperature:
\begin{equation}
\label{eqn:eje_p}
\frac{d p_{\text{ej,p}}}{dt} = \frac{R_{{H}_2} \, T_{\text{ej,p}}}{V_{\text{em}}} \left( W_{\text{HPCV}} - W_{\text{ej,p}} \right),
\end{equation}
\begin{equation}
\label{eqn:eje_T}
\frac{d T_{\text{ej},\text{out}}}{dt} = \frac{T_{\text{ej},\text{out,ss}} - T_{\text{ej},\text{out}}}{\tau_{T,\text{ej}}}.
\end{equation}
Here, Eq.~\eqref{eqn:eje_p}, which results from applying mass conservation to the ejector manifold, also appears as Eq.~(1) in the study by \cite{he2011analysis}. Equation~\eqref{eqn:eje_T} models the non-isothermal dynamics of the ejector outlet using a first-order lag representation, as in earlier sections. As for the states associated with the secondary input of the ejector, the dynamics of $p_{\text{ej,s}}$ and $T_{\text{ej,s}}$ are governed by the upstream anode return manifold model in Section~\ref{sec:an_Manifold}.

\subsubsection{Purge and drain valve}
\label{sec:Purge}
In the anode return manifold, liquid water is periodically drained to prevent accumulation, while purge control is also used to reduce impurity concentrations, particularly nitrogen crossover from the cathode. In the purge and drain valve (PDV) strategy for the FCS, when the valve opens, any accumulated liquid water in the anode return manifold is first drained. Once the liquid is cleared, the valve continues to purge diluent gases, such as nitrogen, that have permeated from the cathode side \cite{falta2015anode}. Based on the described sequential operation of the PDV, the liquid water drain is modeled as a first-order dynamic that drives the water mass to zero:
\begin{equation}
W_{l,\text{drain}}^{\text{an}} = -u_{\text{PDV}} \frac{m_{l,\text{rm}}^{\text{an}}}{\tau_{\text{PDV}}},
\end{equation}
where $u_{\text{PDV}}$ is the control input of the valve. When the liquid water is drained, the gas flow rate  through the valve nozzle, $W_{\text{purge}}$, is determined using a formulation similar to Eq.~\eqref{eqn:wet_outlet_flow}, where the inlet pressure corresponds to the anode return manifold and the outlet pressure is assumed to be atmospheric. The species flow rates of hydrogen, nitrogen, and water vapor through the valve are also calculated based on the total flow rate and the species mass fractions in the return manifold, similar to Eq.~\eqref{eqn:species _frac}. 

\subsubsection{Anode supply and return manifolds}
\label{sec:an_Manifold}
The dynamics of each manifold on the anode side follow the same structure as Eqs.~\eqref{eqn:manifold_pr}--\eqref{eqn:Manifold_T} for the cathode side described in Section~\ref{sec:ca_Manifolds}. The gas flow into the supply manifold is the sum of the primary and secondary ejector flow rates, as computed in Section~\ref{sec:ejector}, whereas the liquid water inlet flow rate is considered zero. On the other hand, the inlet flow rate to the return manifold is determined by the anode outlet of the stack. The outlet flow rates of the manifolds are given by:
\begin{equation}
W^{\mathrm{an}}_{\mathrm{sm,out}} = K^{\mathrm{an}}_{\mathrm{dp,sm}} \left( p^{\mathrm{an}}_{\mathrm{sm}} - p^{\mathrm{an}}_{\mathrm{GC}} \right),
\end{equation}
\begin{equation}
W_{\mathrm{rm,out}}^{\mathrm{an}} =  W_{\mathrm{ej,s}} + W_{\mathrm{purge}}, 
\end{equation}
where \( p_{\mathrm{sm}}^{\mathrm{an}} \) and \( p_{\mathrm{GC}}^{\mathrm{an}} \) denote the total partial pressures of gas species in the anode supply manifold and the anode gas channel of the stack, respectively. In addition, the liquid outlet flow rate from the anode return manifold is equal to \( W_{l,\mathrm{drain}} \), as computed in previous section. 

\subsection{Coolant supply system}
\label{sec:Thermal_management_system}
The coolant supply system, or cooling system, of the FCS dissipates the excess heat generated by electrochemical reactions within the fuel cell stack. It consists of a water pump and an air-cooled radiator coupled with a fan, as shown in Fig.~\ref{fig:FCS}. The input voltages to the pump (\(v_\text{p}\)) and fan (\(v_\text{f}\)) are actively controlled to regulate the coolant flow rate through the radiator and the ambient airflow rate across it. In addition, the maximum power consumption of the fan and pump is assumed to be limited to 1.75~kW and 0.75~kW, respectively. The cooling system model and its integration with the comprehensive FCS model are adopted from our previous work~\cite{ayubirad2024model} and are omitted here for brevity.

\subsection{Stack}
\label{sec:stack}
\begin{figure}[htbp]
\centering
\includegraphics[width=0.7\textwidth]{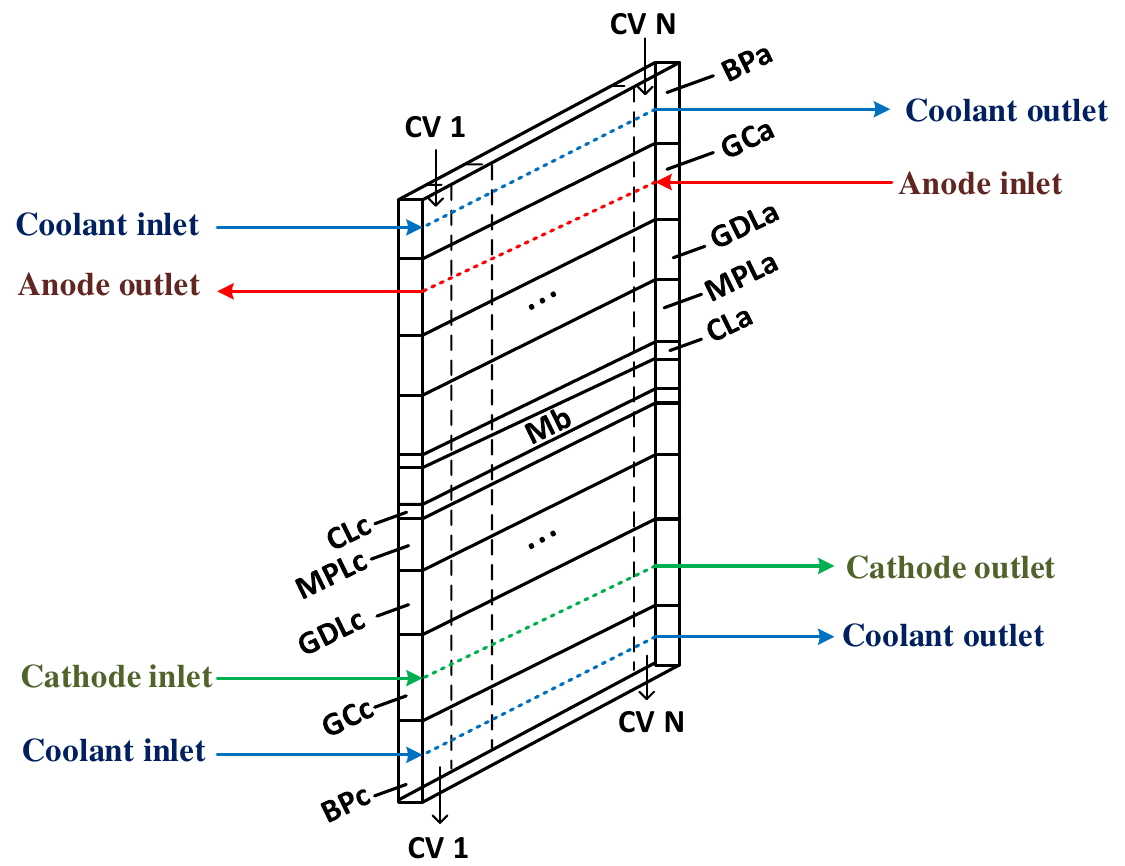}
\caption{Schematic of the pseudo-two-dimensional (P2D) fuel cell model in counter-flow configuration, showing discretization along the channel direction into control volumes (CVs). The layers include: BPa (anode bipolar plate), GCa (anode gas channel), GDLa (anode gas diffusion layer), MPLa (anode microporous layer), CLa (anode catalyst layer), Mb (membrane), CLc (cathode catalyst layer), MPLc (cathode microporous layer), GDLc (cathode gas diffusion layer), GCc (cathode gas channel), and BPc (cathode bipolar plate).}
\label{fig:P2D_cell}
\end{figure}

The one-dimensional (1D) through-the-membrane model of the stack, as presented in the literature \cite{goshtasbi2020degradation,ayubirad2023simultaneous}, can capture the distributions of critical variables through the thickness of the membrane but remains limited in its ability to predict membrane hydration along the gas flow channels. Therefore, it cannot be used to develop control strategies for addressing physical issues such as local flooding or membrane dry-out along the gas channels. To overcome these limitations, this work adopts a P2D stack model, which is obtained by extending the 1D model along the channel, as is commonly done in fuel cell modeling \cite{goshtasbi2020mathematical,kim2010reduced,pant2019along}. Thus, in the P2D model, the same calculations used in the 1D model from our previous publication \cite{ayubirad2023simultaneous} are repeated for each control volume (CV) along the channel, as illustrated in Fig.~\ref{fig:P2D_cell}. The primary distinction between the P2D and 1D models lies in the calculation of local current density at each CV, which varies along the channel due to variations in reactant concentrations along the flow direction. The local current density is obtained by solving the following equation, which is given as Eq.~(52) in Ref.~\cite{yang2019comprehensive}:
\begin{equation}
\left\{
\begin{aligned}
E_{\text{cell},1} &= E_{\text{OCV},1} - \eta_{\text{HOR},1}(i_{\text{dens},1}) - \eta_{\text{ORR},1}(i_{\text{dens},1}) - V_{\text{ohm},1}(i_{\text{dens},1}), \\
&\vdots \\
E_{\text{cell},N} &= E_{\text{OCV},N} - \eta_{\text{HOR},N}(i_{\text{dens},N}) - \eta_{\text{ORR},N}(i_{\text{dens},N}) - V_{\text{ohm},N}(i_{\text{dens},N}), \\
\sum_{i=1}^{N} i_{\text{dens},i} &= N i_{\text{dens}},
\end{aligned}
\right.
\end{equation}
where $\eta_{\text{HOR}}$ and $\eta_{\text{ORR}}$ are the overpotentials for the hydrogen oxidation reaction (HOR) and the oxygen reduction reaction (ORR), respectively; $V_{\text{ohm}}$ is the ohmic overpotential; and $E_{\text{OCV}}$ is the open-circuit voltage of the hydrogen fuel cell, all calculated using Table 2.7 of Ref.~\cite{goshtasbi2019modeling}. Here, the cell potential is assumed to be identical across all CVs, and the average current density, $i_{\text{dens}}$, is assumed to be known a priori as an input to the FCS. The number of CVs, $N$, in the P2D model of the stack is selected to strike a balance between accuracy and computational complexity.

The potential of the P2D model to capture distributions both through the membrane and along the flow channel enables the analysis of spatially variant phenomena such as membrane dry-out and flooding, making it well-suited for developing control strategies for local water management.

\section{Control Development}
\label{sec:Air-supply Subsystem Control Development}
This section presents the development and validation of control strategies for the FCS subsystems, based on the proprietary high-fidelity FCS model provided by Ford Motor Company.

As mentioned previously, the main objectives of the FCS controllers are to deliver maximum net power from the fuel cell while protecting the stack from damages caused by oxygen starvation and membrane dry-out or flooding. Accordingly, the following sections begin by identifying the optimal set-points for the air supply and cooling subsystems through steady-state power optimization in which constraints are satisfied. Next, controller development is presented to compute the current demand required to meet the requested power, as well as the control inputs, namely the compressor speed reference and the throttle valve position for the air supply system, the HPCV and purge valve commands for the fuel delivery system, and the coolant pump and radiator fan voltages for the cooling system, as illustrated in Fig.~\ref{fig:FCS}. 

\begin{figure}[htbp]
\centering
\includegraphics[width=0.65\textwidth]{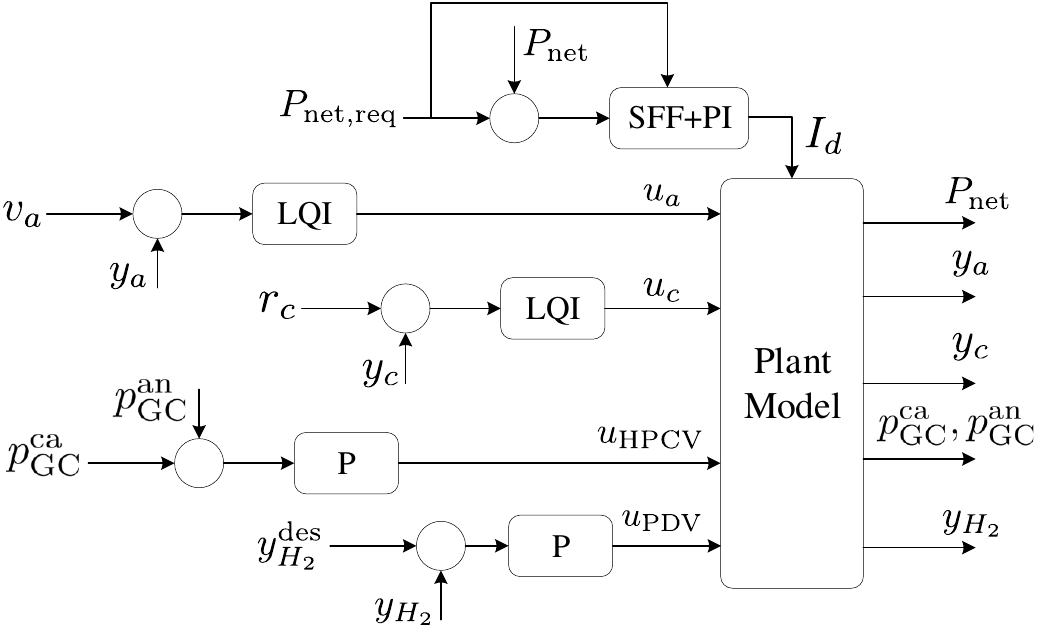}
\caption{ Block diagram of the control architecture for the automotive FCS.}
\label{fig:control_scheme}
\end{figure}

The layout of the controllers used in this work is illustrated in Fig.~\ref{fig:control_scheme}. The control architecture shown in Fig.~\ref{fig:control_scheme} operates under the constraint-aware set-point management structure presented in Fig.~\ref{fig:total_scheme}. Specifically, the input \( v_\text{a} \) is obtained by processing the optimal set-point \( r_\text{a} \) through the CG-based constraint management in Fig.~\ref{fig:total_scheme} to produce a constraint-aware command that serves as the reference for the air supply controller. This reference is tracked by the LQI controller, which corresponds to the ``Air Supply Controller'' block shown in Fig.~\ref{fig:total_scheme}.
In parallel, the optimal set-point for the cooling system, \( r_\text{c} \), is generated alongside \( r_\text{a} \) but is not modified by the CG-based constraint management. Instead, \( r_\text{c} \) is passed directly to the LQI controller, which corresponds to the ``Cooling System Controller'' block in Fig.~\ref{fig:total_scheme}. The fuel delivery and power tracking controllers are only included in Fig.~\ref{fig:control_scheme}, as they do not play a direct role in constraint management. 

As shown in Fig.~\ref{fig:control_scheme}, the air and coolant supply subsystems are regulated using LQI controllers, while proportional (P) controllers are implemented on the anode side to compute the HPCV and PDV commands. For current demand computation, a combination of static feedforward (SFF) and proportional–integral (PI) control is employed. In this control structure, the coolant supply control is decoupled from the air supply and fuel delivery controls, as temperature evolves slowly and has negligible influence on the fast dynamics of the other subsystems. The air supply and fuel delivery controllers are also designed independently, as the control actions of each subsystem have minimal cross-coupling effects on the others. Further details about the controller design are provided in the following subsections.

\subsection{Set-point maps}
\label{sec:setpoint_map}
In this subsection, the high-fidelity model is used to generate the optimal set-points for the air and coolant supply subsystems. We assume that the ambient conditions, including temperature, pressure, and relative humidity remain constant. As a result, the optimal set-points for the air and coolant supply subsystems are functions of the current demand only\footnote{If these assumptions are not satisfied, the optimal set-points for the FCS can instead be defined as functions of both current demand and ambient conditions.}, as illustrated in Fig.~\ref{fig:total_scheme}. It is also assumed that the auxiliary subsystems, including the compressor motor, coolant pump, and radiator fan, are powered by the FCS itself. Therefore, the net power output of the system is given by:
\begin{equation}
\label{eqn:P_net}
P_{\text{net}}(t) = P_{\text{st}}(t) - P_{\text{cm}}(t) - P_{\text{f}}(t) - P_{\text{p}}(t),
\end{equation}
where \( P_{\text{st}} \) is the power generated by the fuel cell stack, and \( P_{\text{cm}} \), \( P_{\text{f}} \), and \( P_{\text{p}} \) denote the electrical power consumption of the compressor, radiator fan, and coolant pump, all driven by electric motors. To determine the optimal set-points corresponding to each current demand \( I_{\text{d}} \), the net power output is maximized at steady-state, with all system constraints satisfied, using an optimization procedure similar to that described in Section~3.1 of our previous publication~\cite{ayubirad2023simultaneous}.
The applied constraints during power optimization at steady-state include a lower bound on oxygen excess ratio (OER), upper and lower limits on membrane hydration, and linear inequalities that prevent compressor surge and chokes. The OER is defined as the ratio between the oxygen supplied $W_{\text{O}_2, \text{in}}^{\text{ca}}$ and the oxygen consumed $W_{\text{O}_2, \text{rxn}}^{\text{ca}}$ in the cathode:
\begin{equation}
\lambda_{\text{O}_2} = \frac{W_{\text{O}_2, \text{in}}^{\text{ca}}}{W_{\text{O}_2, \text{rxn}}^{\text{ca}}}
\end{equation}
To prevent oxygen starvation, a lower limit of $ \lambda_{\text{O}_2} \ge 1.6$ is imposed. Similarly, hydration is maintained by constraining the membrane water content \( \lambda_{\mathrm{mb}} \) to remain between a lower bound of \( \lambda_{\mathrm{mb,lb}} = 6\;[-] \) and an upper bound of \( \lambda_{\mathrm{mb,ub}} = 14\;[-] \)\footnote{The determination of safety bounds remains an active area of research. The limits adopted in this work are used solely to demonstrate the effectiveness of the proposed control strategy and should not be regarded as general safety thresholds.}. The dynamics of membrane water content are governed by:
\begin{equation}
\frac{d \lambda_{\text{mb}}}{dt} = \frac{EW_{}}{\rho_{\text{ion}} \, \delta_{\text{mb}}} \left( N_{w}^{\text{an2mb}} - N_{w}^{\text{mb2ca}} \right),
\end{equation}
where $N_{w}^{\text{an2mb}}$ and $N_{w}^{\text{mb2ca}}$ represent the water flux in the ionomer phase from anode catalyst layer to  membrane and from membrane to cathode catalyst layer, respectively~\cite{goshtasbi2019modeling}.
Additionally, for the safe operation of the compressor, the surge and choke constraints are defined by the following linear inequalities:
\begin{equation}
\frac{p_{\mathrm{sm}}^\mathrm{ca}}{p_{\mathrm{atm}}} \leq 1 + 0.005W_{\mathrm{cp}}, \quad
\frac{p_{\mathrm{sm}}^\mathrm{ca}}{p_{\mathrm{atm}}} \geq 1 + 0.00125W_{\mathrm{cp}}
\end{equation}

The set-point map for the air and coolant supply systems, which satisfy the above constraints at steady-state, are adopted from our previous publications \cite{ayubirad2023simultaneous,ayubirad2024model} as representative examples of optimal set-points and are not reproduced here for brevity.

\subsection{Fuel delivery system control}
\label{sub:FDS_control}
The fuel delivery subsystem features two actuators: an HPCV, which regulates the inlet hydrogen flow rate, and a PDV, which controls water drainage and anode gas exhaust to the ambient, as shown in Fig.~\ref{fig:FCS}. Since a high pressure difference between the cathode and anode can potentially damage the thin polymer electrolyte membrane, the HPCV is used to adjust the hydrogen flow rate such that the anode pressure closely follows any changes in cathode pressure. The HPCV controller adopts a proportional control structure, similar to that used in~\cite{pukrushpan2004control3}, and is given by:
\begin{equation}
u_{\text{HPCV}} = K_{\text{HPCV}} (p^{\text{ca}}_{\text{GC}} - p^{\text{an}}_{\text{GC}}),
\end{equation}
where $K_{\text{HPCV}}$ is the constant gain, manually tuned via simulation to ensure that the anode pressure $p^{\text{an}}_{\text{GC}}$ closely tracks the cathode pressure $p^{\text{ca}}_{\text{GC}}$, without introducing oscillations or overshoot in the anode pressure response.

To remove impurities and maintain optimal gas composition in the anode loop, the PDV is assumed to regulate the hydrogen mole fraction in the anode return manifold to an optimal value \( y_{{H}_2}^{\mathrm{des}} \), using a proportional controller:
\begin{equation}
u_{\mathrm{PDV}} = K_{\mathrm{PDV}} \left( y_{{H}_2}^{\mathrm{des}} - y_{{H}_2} \right),
\end{equation}
where \( K_{\mathrm{PDV}} \) is the proportional gain and is tuned following the same procedure used for $K_{\text{HPCV}}$. The PDV control of the FCS is a more challenging problem, which has been addressed in many publications \cite{he2011analysis,nikiforow2013optimization}. The simplification of PDV control in this paper aims to reduce the complexity of the simulation and allows us to focus on controlling the membrane hydration, which is the focus of this study. This simplification is further justified by previous findings showing that anode purging has a negligible effect on membrane hydration \cite{goshtasbi2021model}.

\subsection{Air supply system control}
\label{sub:Air_supply_control} 
This section focuses on the control of the air supply subsystem. First, reduced-order linear models are extracted from the high-fidelity model at multiple operating points to capture the system’s behavior under varying conditions. Next, the system’s nonlinearity and input–output coupling are examined across the operating range. Based on these characteristics, an LQI controller is designed to track the optimal set-points corresponding to steady-state power optimization, as described in Section~\ref{sec:setpoint_map}.

\subsubsection{Model identification and analysis}
\begin{figure}
    \hspace*{-1em}
    \subfloat[]{%
    \includegraphics[width=0.55\textwidth]{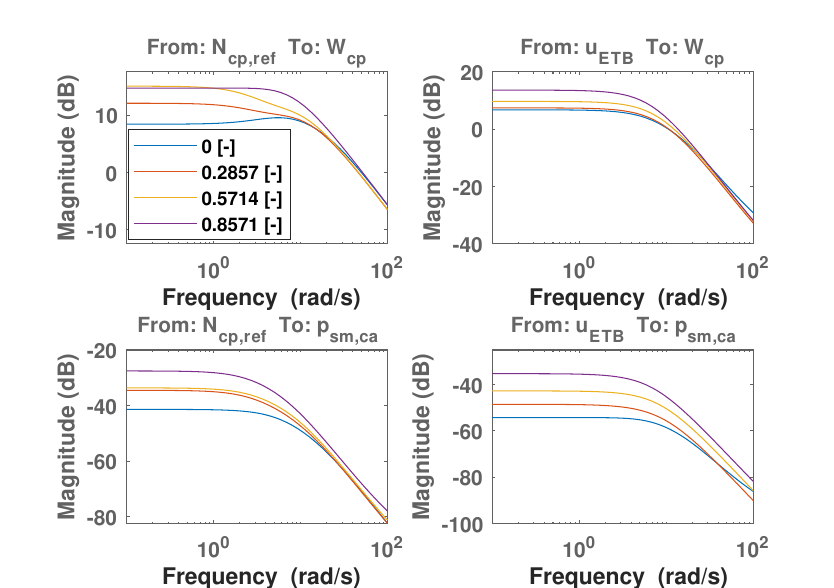}
        \label{fig:Bode_gain}
    }\hspace*{-2.3em}%
    \subfloat[]{%
       \includegraphics[width=0.55\textwidth]{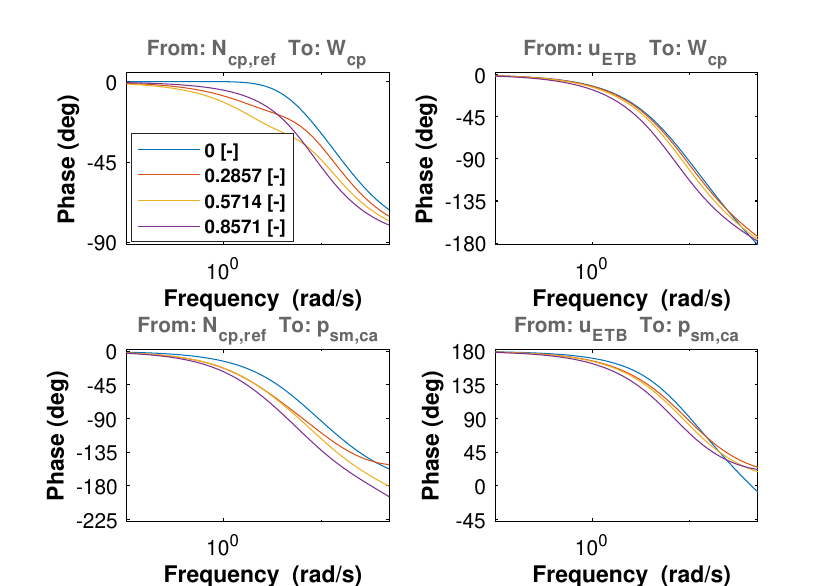}
        \label{fig:Bode_phase}
    }
    \caption{Bode plots of the air supply system’s transfer functions from compressor speed 
and ETB opening to compressor flow rate and cathode supply manifold pressure for various stack currents,: (a) magnitude and (b) phase. The notations \( p_{\mathrm{sm,ca}} \) and \( p_{\mathrm{sm}}^{\mathrm{ca}} \) are used interchangeably to denote the same physical quantity, the pressure at the cathode-side supply manifold.
}
    \label{fig:Bode}
\end{figure}

The high-fidelity FCS model is a high-order system with over 300 states. We follow the same model reduction strategy and the nonlinearity and coupling analysis procedure described in our previous publications \cite{ayubirad2023simultaneous, bacher2023hierarchical}; therefore, the details are omitted here for brevity. To capture the system’s behavior under different load conditions, we derive reduced-order linear models at four operating points corresponding to normalized stack currents of 0, 0.2857, 0.5714, and 0.8571. Variables that are scaled in this study are presented in this form to protect proprietary details of the high-fidelity system provided by Ford Motor Company. Each model is represented by a linear time-invariant (LTI) state-space system:
\begin{equation}
\begin{aligned}
\delta \dot{x} &= A \, \delta x + B \, \delta u_a, \\
\delta y_a &= C \, \delta x,
\end{aligned}
\label{eqn:lin_model}
\end{equation}
where $\delta(\cdot) = (\cdot) - (\cdot)_{\text{eq}}$ denotes the deviation from the equilibrium value. For simplicity, the $\delta$ notation is omitted for the remainder of the paper.   As a representative case, the state-space matrices for the operating point at a normalized current of 0.2857~[--] are presented here: 
\begin{equation}
\begin{aligned}
A &= \begin{bmatrix}
-14.5 & -1.51 & 2.04 \\
-2.864 & -0.598 & -8.279 \\
-18.52 & -0.2867 & -6.701
\end{bmatrix}, \quad
B = \begin{bmatrix}
-1.506 & 0.6151 \\
7.274 & -0.4801 \\
0.2365 & 0.5155
\end{bmatrix}, \\[1.5ex]
C &= \begin{bmatrix}
55.57 & 19.71 & -50.04 \\
-0.0539 & -0.01246 & 0.05193
\end{bmatrix}.
\end{aligned}
\label{eqn:ABC_matrices}
\end{equation}
To evaluate the degree of nonlinearity, Bode plots for the four operating points are shown in Fig.~\ref{fig:Bode}. While a full nonlinearity analysis is omitted, the key observation is that the airflow exhibits moderate nonlinearity, whereas the pressure response demonstrates a higher degree of nonlinearity. Nevertheless, a single linear controller can achieve satisfactory performance across the full operating range.
As for the coupling analysis, the findings are consistent with our previous publication \cite{bacher2023hierarchical} and indicate that to achieve high performance in the presence of strong input–output coupling in the air supply dynamics, a multi-input multi-output (MIMO) controller is recommended.

\subsubsection{Controller design and validation}
\label{sub:controller_design} 
In this section, we design a MIMO controller for the air supply system using the LQI approach. While an MIMO IMC strategy was successfully applied in our previous works \cite{ayubirad2023simultaneous, bacher2023hierarchical}, its application to the high-fidelity air supply model in this study proved challenging. Specifically, the IMC parameters could not be tuned to achieve satisfactory tracking performance without inducing aggressive control inputs. To address this limitation, the LQI approach was selected instead, as it offers more tuning flexibility and enables the desired tracking performance while mitigating control input aggressiveness.

To design the LQI controller, the state vector is first augmented with the integrals of the output tracking errors, yielding the augmented model:
\begin{equation}
\begin{aligned}
\begin{bmatrix}
\dot{x} \\
\dot{q}
\end{bmatrix}
&=
\begin{bmatrix}
A & 0 \\
C & 0
\end{bmatrix}
\begin{bmatrix}
x \\
q
\end{bmatrix}
+
\begin{bmatrix}
B \\
0
\end{bmatrix}
u_a
+
\begin{bmatrix}
0 \\
-I
\end{bmatrix}
v_a, \\[1ex]
y_a &= 
\begin{bmatrix}
C & 0
\end{bmatrix}
\begin{bmatrix}
x \\
q
\end{bmatrix},
\end{aligned}
\label{eq:augmented_ss}
\end{equation}
where
\begin{equation*}
\dot{q} = 
\begin{bmatrix}
W_{\text{cp}} - W_{\text{cp}}^{\text{mod}} \\
p_{\text{sm,ca}} - p_{\text{sm,ca}}^{\text{mod}}
\end{bmatrix}.
\end{equation*}
The LQI feedback gains are obtained by minimizing the following cost function:
\begin{equation}
J = \int_{0}^{\infty} y_a^\top Q_y y_a + q^\top Q_I q + {u_a}^\top R u_a \, dt,
\label{eq:cost_function}
\end{equation}
where \( Q_y, Q_I, \) and \( R \) are the weighting matrices on the output \( y_a \), integrator state \( q \), and control input \( u_a \).  To minimize this cost, the following state feedback control law is used:

\begin{equation}
u_a = 
\begin{bmatrix}
- K_{p,\mathrm{LQI}} & - K_{i,\mathrm{LQI}}
\end{bmatrix}
\begin{bmatrix}
x \\
q
\end{bmatrix}
\label{eq:control_law},
\end{equation}
where the gains $K_{p,\mathrm{LQI}}$ and $K_{i,\mathrm{LQI}}$ are computed by solving the Algebraic Riccati Equation. 

The controller parameters are tuned using the linearized air supply model evaluated at the normalized stack current of 0.2857~[--], as given in equations  \eqref{eqn:lin_model}--\eqref{eqn:ABC_matrices}. The weighting matrices below are chosen to achieve an appropriate trade-off between output tracking and input effort:
\begin{equation}
Q_y = \begin{bmatrix} 80 & 0 \\ 0 & 75000 \end{bmatrix}, \quad
Q_I = \begin{bmatrix} 2500 & 0 \\ 0 & 200000 \end{bmatrix}, \quad
R = \begin{bmatrix} 10 & 0 \\ 0 & 0.9 \end{bmatrix},
\label{eq:lqi_weights}
\end{equation}
which result in the following LQI gains:
\begin{equation}
K_{p,\mathrm{LQI}} = 
\begin{bmatrix}
163.1506 & 55.9856 & -140.5674 \\
75.3874 & 18.6795 & -46.9910
\end{bmatrix}, \quad
K_{i,\mathrm{LQI}} = 
\begin{bmatrix}
15.7474 & 12.7071 \\
4.7357 & -469.4977
\end{bmatrix}
\label{eq:lqi_gains}.
\end{equation}

The LQI controller requires all the states of the system to be either measured or estimated. However, the states of the reduced-order model lack physical meaning and cannot be directly measured. Therefore, similar to our previous work \cite{ayubirad2023simultaneous}, a steady-state Kalman filter is used to estimate the system states based on the available measurements \( y_a = \begin{bmatrix} W_{\text{cp}}, p_{\text{sm,ca}} \end{bmatrix}^\top \). 

To evaluate the controller’s robustness, we quantify the stability margins of the closed-loop system under gain and phase variations using disk-based gain margin (DGM) and disk-based phase margin (DPM). Specifically, we consider multiloop uncertainty applied independently at the plant inputs, at the outputs, and simultaneously at both. These margins provide a measure of how much gain or phase deviation the system can tolerate before losing stability.  Table~\ref{tab:stability_margins} reports the stability margins at the operating conditions corresponding to the scaled stack current values mentioned previously. The robustness analysis indicates that the feedback controller exhibits high multiloop robustness to gain and phase variations applied at the inputs or outputs, and satisfactory robustness when such variations are applied simultaneously at both, across the full range of operating conditions. Thus, the linear controller, designed at the nominal operating point, can be applied across the entire operating range with  closed-loop stability and acceptable robustness.
\begin{table}[t!]
\centering
\renewcommand{\arraystretch}{1.2}
\begin{tabular}{c c c c c c c}
\hline
Scaled current [-] & DGM\textsubscript{i} & DPM\textsubscript{i} & DGM\textsubscript{o} & DPM\textsubscript{o} & DGM\textsubscript{io} & DPM\textsubscript{io} \\
\hline
0       & $\pm$11.61 dB & 60.58$^{\circ}$ & $\pm$11.46 dB & 76.38$^{\circ}$ & $\pm$6.15 dB & 37.53$^{\circ}$ \\
0.2875  & $\pm$11.04 dB & 58.67$^{\circ}$ & $\pm$16.07 dB & 72.13$^{\circ}$ & $\pm$5.98 dB & 36.64$^{\circ}$ \\
0.5741  & $\pm$11.83 dB & 61.26$^{\circ}$ & $\pm$17.18 dB & 74.25$^{\circ}$ & $\pm$6.15 dB & 37.56$^{\circ}$ \\
0.8571  & $\pm$10.96 dB & 58.41$^{\circ}$ & $\pm$17.03 dB & 73.98$^{\circ}$ & $\pm$5.68 dB & 35.05$^{\circ}$ \\
\hline
\end{tabular}
\caption{\centering Disk-based gain and phase margins at selected operating points. Here, the same controller, designed at the normalized current of 0.2875~[--], is evaluated across all operating conditions.}
\label{tab:stability_margins}
\end{table}

\begin{figure}[t!]
\centering
\subfloat{%
  \includegraphics[width=0.55\textwidth]{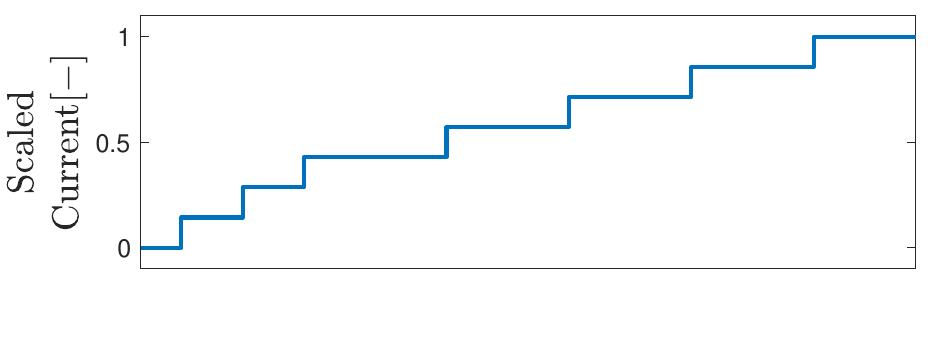}
}
\vspace{-2.5em}
\subfloat{%
  \includegraphics[width=0.55\textwidth]{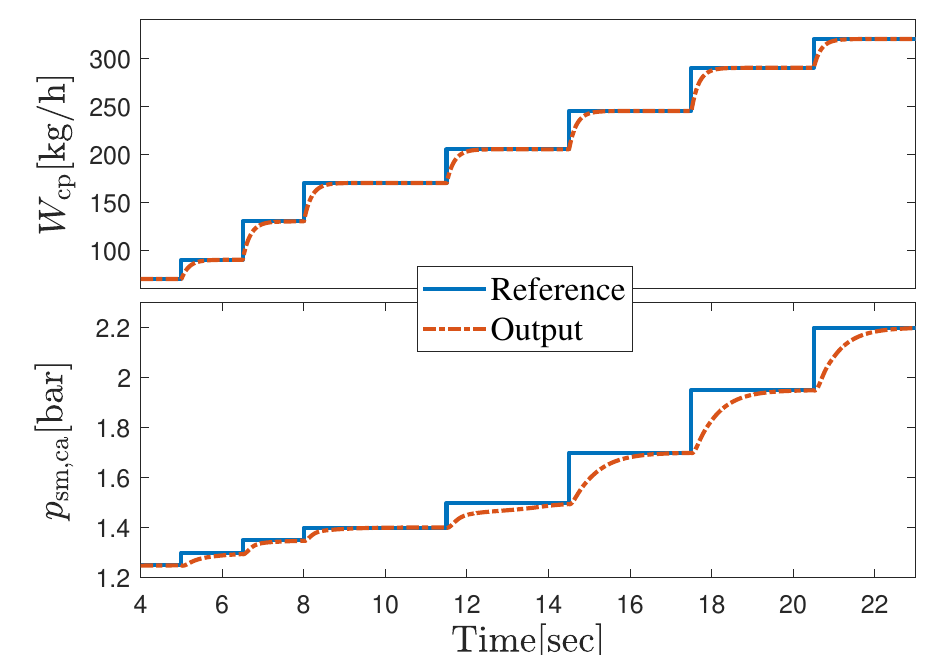}
}
\caption{Response of the air supply system to a sequence of step changes in stack current, simulated using the high-fidelity model with the LQI controller.}
\label{fig:air_path_response}
\end{figure}

To evaluate how well the nominal LQI controller performs across different operating conditions of the nonlinear system, it is applied to the high-fidelity model using a sequence of stack current steps. The corresponding current demand, compressor mass flow rate, and cathode inlet pressure responses are presented in Fig.~\ref{fig:air_path_response}. As illustrated in the figure, the compressor mass flow rate exhibits similar dynamic behavior across the operating range, as expected given the channel’s previously described moderate nonlinearity. In contrast, the inlet pressure dynamics of the cathode vary between low and high load conditions, reflecting the higher degree of nonlinearity in this channel, as identified earlier. Nonetheless, applying the same LQI formulation across all operating points yields satisfactory performance, with minimal overshoot and zero steady-state tracking error, thus eliminating the need for gain scheduling.

Moreover, the distinct temporal characteristics of pressure and air flow rate observed in Fig.~\ref{fig:air_path_response} are consistent with the closed-loop bandwidths achieved under the LQI controller, which are approximately 5.5 [rad/s] for the compressor mass flow rate and 1.4 [rad/s] for the cathode inlet pressure. This bandwidth separation helps decouple the control loops, thereby mitigating performance degradation due to dynamic interactions.

\subsection{Cooling system control}
\label{sub:Temperature_control} 
For temperature control, we adopt the LQI controller presented in our previous publication \cite{ayubirad2024model}. This choice is motivated by its ability to achieve satisfactory tracking performance without inducing aggressive control inputs, as was also the case for the air supply system in previous section. The controller regulates both the coolant inlet ($T_{\mathrm{c,in}}$) and outlet ($T_{\mathrm{c,out}}$) temperatures of the stack. Specifically, $T_{\mathrm{c,out}}$ is targeted because, in a counter-flow configuration, it represents the temperature at the driest region of the membrane, making it critical for water management. Meanwhile, $T_{\mathrm{c,in}}$ is regulated to control the temperature gradient along the flow direction, thereby mitigating thermal stresses. The controller ensures zero steady-state error with minimal to zero overshoots and was developed using the same high-fidelity model employed in the present study\footnote{The model was not presented in full in \cite{ayubirad2023simultaneous}, and only key components relevant to controller design were included.}. For brevity, the full derivation is omitted here; the reader is referred to \cite{ayubirad2024model} for details on model identification, analysis, and controller design for cooling system.

\subsection{Power Tracking Control}
\label{sub:Power_control} 
\begin{figure}[htbp]
\centering
\includegraphics[width=0.45\textwidth]{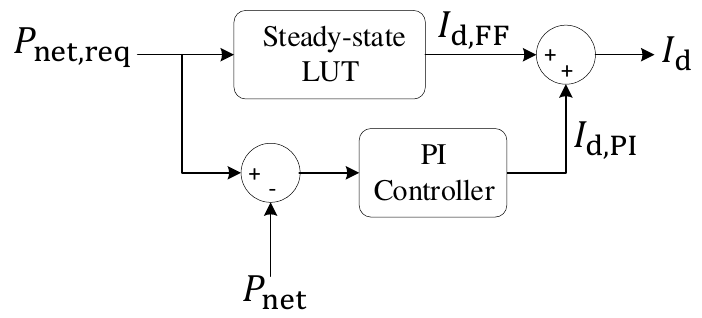}
\caption{Block diagram of the proposed power tracking controller.}
\label{fig:Power_controller}
\end{figure}

For power tracking control, we consider a two-degrees-of-freedom controller based on feedforward and feedback, as shown in Fig.~\ref{fig:Power_controller}. The feedforward path employs a steady-state map implemented via a lookup table (LUT), which translates the requested net power into a corresponding current demand, $I_{d,\mathrm{FF}}$. This map is generated by simulating the high-fidelity model, or experimentally, if a hardware platform is available, with closed-loop air supply and cooling system over a range of current values. The resulting steady-state difference between the fuel cell stack output power and the compressor power consumption is then recorded. The feedforward current is subsequently combined with a slow PI feedback controller:
\begin{equation}
I_{\mathrm{d,PI}} = K_p \left(P_{\text{net,req}} - P_{\text{net}}\right) + K_i \int \left(P_{\text{net,req}} - P_{\text{net}}\right)\, dt
\label{eqn:Power_PI}
\end{equation} 
to determine the final current demand, $I_\mathrm{d}$, required to deliver the requested net power. This configuration leverages the disparate timescales of the power terms in Eq.~\ref{eqn:P_net}: the LUT accounts for the fast dynamics of $P_{\mathrm{st}}$ and $P_{\mathrm{cm}}$, while the PI controller compensates for the slower dynamics of $P_{\mathrm{f}}$ and $P_{\mathrm{p}}$. The PI gains are tuned using the high-fidelity model. A small $K_i$ is selected to reflect the slow dynamics of power consumption in the cooling system, without making the response sluggish, while $K_p$ is set high enough to improve the tracking performance without inducing aggressive current changes.

\section{Hydration constraint management}
\label{sec:Hydration_constraint_management}

\subsection{Problem description}
As mentioned previously, studies such as~\cite{goshtasbi2021model} have shown that under counter-flow operation of the stack, the driest region of the membrane tends to form near the anode inlet, whose location within the stack is shown in Fig.~\ref{fig:P2D_cell}. In contrast, the cathode inlet and the central portion of the membrane tend to be more hydrated. More specifically, the membrane water content in the $N$-th CV (see Fig.~\ref{fig:P2D_cell}), exhibits the minimum value across all segments. This is primarily because the anode inlet coincides with the coolant outlet, making it the hottest region of the stack. The low vapor partial pressure at this location, combined with minimum water generation due to the low local current density, further contributes to the reduced membrane hydration in this region. 
\begin{figure}[b!]
\centering
\includegraphics[width=0.5\textwidth]{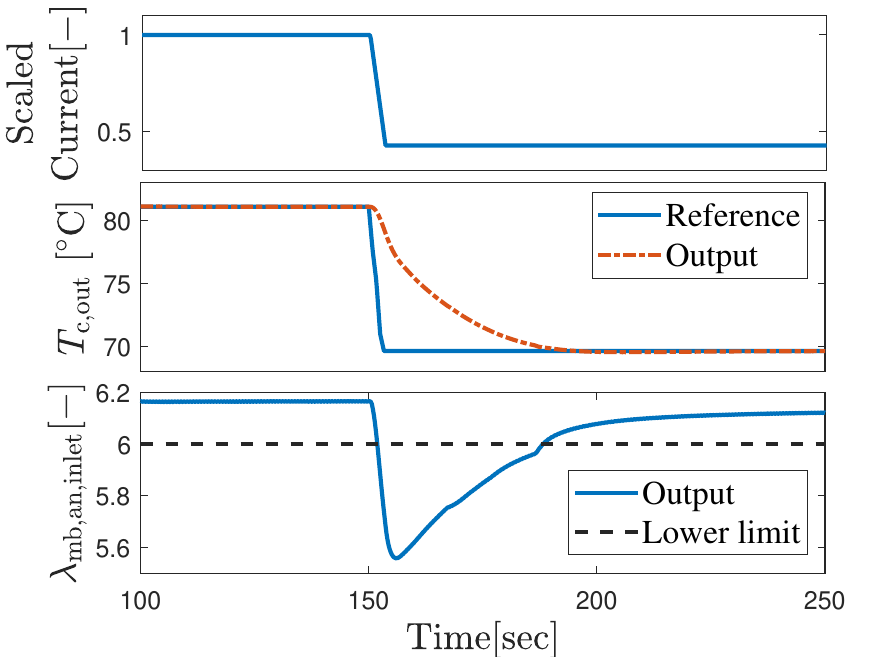}
\caption{Response of the fuel cell system under a load reduction following high-temperature operation.}
\label{fig:hydration_transition}
\end{figure}

Fig.~\ref{fig:hydration_transition} shows the membrane water content at the anode inlet, denoted as $\lambda_{\mathrm{mb},,\mathrm{an,inlet}}$, during a load reduction following a high-temperature condition caused by sustained high- load operation. It can be seen that the membrane water content at the anode inlet, as the driest region of the stack, violates the lower hydration bound at transient. The simulation results are consistent with those reported~\cite{kitamura2010development}. While dry-out is a localized phenomenon within the membrane, this localized failure can propagate and lead to failure of the entire stack.

As illustrated in Fig.~\ref{fig:hydration_transition}, the hydration constraint violation during transients is caused by the inherently slow response of the thermal system, which cannot reduce the coolant outlet temperature fast enough. It is worth noting that the cooling system set-points were previously designed to ensure no constraint violations at steady-state; thus, the cooling system does contribute to hydration management, but only on a slower timescale.  Therefore, this paper explores the use of air supply system, characterized by its faster dynamic response, as a means of mitigating constraint violations during transient operation. To this end, the high-fidelity closed-loop model of the FCS is employed to design a predictive control strategy for membrane hydration management at the anode inlet. The predictive constraint enforcement scheme is formulated within the CG framework, which ensures constraint satisfaction by temporarily modifying the air supply set-points when necessary.


\subsection{Control-oriented model}
\label{sec:Control-oriented model}
The large-scale nonlinear high-fidelity model described previously is computationally intractable for real-time optimization-based constraint enforcement. One possible way of reducing the computational burden is to use a low-order linear model of the closed-loop FCS system for the construction of the governor scheme.

To ensure that the linear model captures the nonlinear system behavior around the high-load operating condition, where the stack is most prone to dry-out, the closed-loop FCS model is linearized around an operating point corresponding to a normalized stack current of 0.9~[--]. 
To obtain a control-oriented approximation, the balanced truncation technique \cite{brunton2022data} is applied to the full-order linear model to eliminate states with limited influence on the hydration dynamics at the anode inlet. Following the truncation, an eighth-order model is selected to preserve the time-domain behavior of the full-order system. The reduced eight-state model is provided in Appendix A. Fig.~\ref{fig:hydration_comparison} compares the hydration response of the reduced-order  model to that of the nonlinear high-fidelity model under varying operating conditions. As illustrated in the figure, the reduced-order linear model accurately captures the hydration dynamics  across medium to high load conditions. At lower loads, it underestimates the membrane water content, which may lead to a more conservative enforcement of the hydration constraint. This conservative behavior, together with the model’s simplicity and sufficient fidelity, supports the use of a single linear approximation for the design of governor scheme for hydration constraint enforcement. In the following subsection, we first provide a review of the CG scheme before introducing the modification of CG employed in this study for hydration constraint enforcement.
\begin{figure}
\centering
\includegraphics[width=0.55\textwidth]{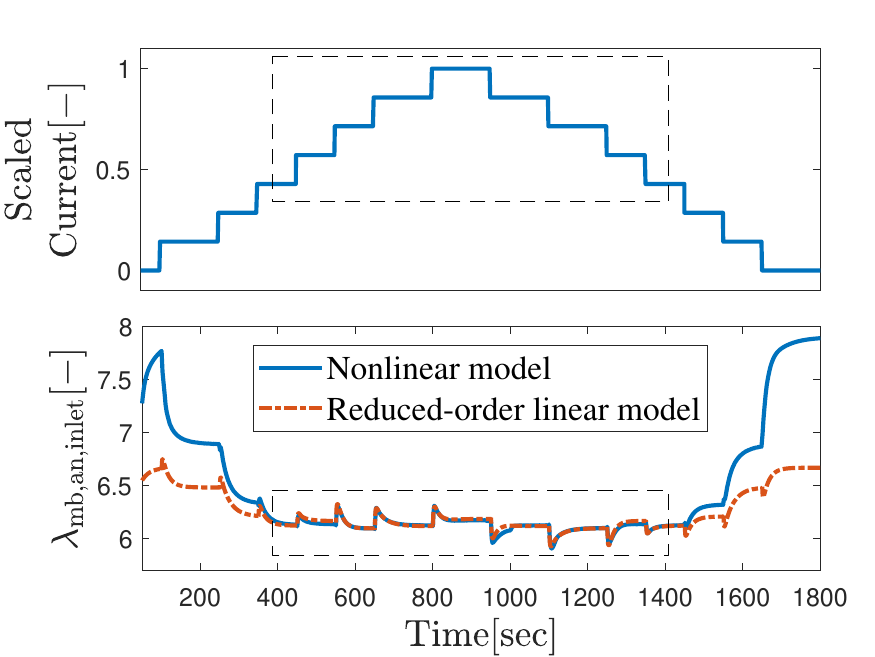}
\caption{Hydration response of the closed-loop FCS. The dashed black box highlights the operating region where the response of the nonlinear system closely matches that of the reduced-order linear model.}
\label{fig:hydration_comparison}
\end{figure}
\subsection{Review of command governors}
\label{sec:CG_review}
\begin{figure}[b!]
\centering
\includegraphics[width=0.6\textwidth]{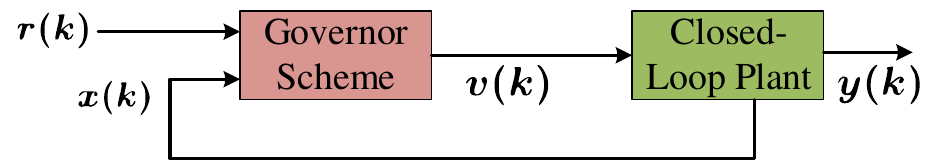}
\caption{Reference governor controller scheme. The signals are as follows: $y(k)$ is the constrained output, $r(k)$ is the desired reference, $v(k)$ is the modified reference command, and $x(k)$ is the system state.}
\label{fig:RG_scheme}
\end{figure}

Consider Fig.~\ref{fig:RG_scheme}, where the “Closed-loop Plant” denotes a discrete-time LTI system in feedback with a stabilizing controller, with closed-loop dynamics described by:

\begin{equation}
\begin{aligned}
x(k+1) &= A_{\text{cl}}\,x(k) + B_v\,v(k), \\
y(k) &= C_{\text{cl}}\,x(k) + D_v\,v(k),
\end{aligned}
\label{eqn:discrete_sys}
\end{equation}
where $x(k)\in {\mathbb{R}}^{n}$ is the state vector, $v(k)\in \mathbb{R}^{m}$ is the input vector, and \mbox{$y(k)\in {\mathbb{R}}$} is the constrained output. Constraints are imposed on the output variables as \mbox{$y(k) \leq s$} for all $k \in \mathbb{Z}_+$, where $\mathbb{Z}_+$ denotes the set of all nonnegative integers. The CG approach computes a modified reference command which, if held constant, guarantees constraint satisfaction for all future times. To achieve this, the CG leverages an inner approximation of the maximal admissible set~\cite{garone2017reference}, defined as the set of all initial conditions and constant inputs such that the ensuing output satisfies the constraints for all time:
\begin{equation}
\Omega = \left\{ (x_0, v_0) : x(0) = x_0,\ v(k) = v_0,\ \Rightarrow\ y(\infty) \leq (1-\epsilon)s,\ y(k) \leq s,\ k = 0, \ldots, k^* \right\},
\label{eqn:omega}
\end{equation}
where a tightened steady-state constraint, $y(\infty) \leq (1-\epsilon)s$ with $0 < \epsilon \ll 1$, together with mild assumptions on $C_{\mathrm{cl}}$ and $A_{\mathrm{cl}}$, ensures that the set $\Omega$ is finitely determined
~\cite{gilbert1991linear}.
Specifically, there exists an admissibility index $k^*$ such that if the output constraints are satisfied for $k = 0, \dots, k^*$, they are guaranteed to be satisfied for all $k > k^*$.
For the LTI system \eqref{eqn:discrete_sys}, $\Omega$ can be explicitly characterized as:
\begin{flalign}
\Omega = \left\{ (x_0, v_0) \;\middle|\; H_x x_0 + H_v v_0 \leq h \right\}
\label{eqn:omega_e}
\end{flalign}
where
\[
H_x = \begin{bmatrix}
0 \\ 
C_{\text{cl}} \\ 
C_{\text{cl}} A_{\text{cl}} \\ 
\vdots  \\ 
C_{\text{cl}} A_{\text{cl}}^{k^*}
\end{bmatrix}, \quad
H_v = \begin{bmatrix}
C_{\text{cl}}(I - A_{\text{cl}})^{-1} B_v + D_v \\
D_v \\
C_{\text{cl}}(I - A_{\text{cl}})(I - A_{\text{cl}})^{-1} B_v + D_v \\
\vdots \\
C_{\text{cl}}(I - A_{\text{cl}}^{k^*})(I - A_{\text{cl}})^{-1} B_v + D_v
\end{bmatrix}, \quad
h = \begin{bmatrix}
(1 - \epsilon)s \\
s \\
\vdots \\
s
\end{bmatrix}.
\]
The CG directly optimizes over \( v(k) \) by solving the following QP at every timestep:
\begin{equation}
\begin{aligned}
v(k) = \arg\min_{v} \ & \|v - r(k)\|^2  \\
\text{s.t. } \ & (x(k),v)\in \Omega ,
\end{aligned}
\label{eqn:QP_problem}
\end{equation}
where $x(k)$ is the currently available state at time step $k$. The QP problem in \eqref{eqn:QP_problem} is based on the idea of computing, at each time step $k$ and on the basis of the current state $x(k)$, the optimal control input $v(k)$ that is as close as possible to $r(k)$ while ensuring constraint satisfaction, i.e., $(x(k), v(k)) \in \Omega$. 

For comparison of CG with the conventional RG in the scalar case, the CG is computationally more demanding than the RG~\cite{garone2017reference}.
The CG, however, offers greater capability, as it can slow down the command or introduce the minimum necessary overshoot of the command to satisfy the constraints, which, as we will show, is necessary for effective constraint enforcement of localized hydration without degrading power tracking performance. The RG, by contrast, can only slow the desired command by ensuring that \( v(k) \) remains within the interval defined by \( v(k{-}1) \) and \( r(k) \), which is not suitable for hydration management.

\subsection{Hydration management command governor}
\label{sec:hydration_RG}
In this section, the goal is to augment the closed-loop FCS with the CG-based constraint management (shown in Fig.~\ref{fig:total_scheme}) to minimize the gap between ${v_\text{a}}$ and ${r_\text{a}}$ while satisfying the constraints, as in the CG review section. As mentioned previously, the modified command vector is defined as \( {v}_\text{a} = [W^{\text{mod}}_{\mathrm{cp}},\; p^{\text{mod}}_{\mathrm{sm,ca}}]^\top \), representing the constraint-satisfying version of the desired air supply command vector \( {r}_\text{a} = [W^{\text{des}}_{\mathrm{cp}},\; p^{\text{des}}_{\mathrm{sm,ca}}]^\top \). Meanwhile, the current demand \( I_{\text{d}} \) and the  optimal set-points of the cooling system \( r_\text{c} \) are not modified and are applied directly. 

To design the CG controller for hydration management, it is first necessary to understand how the set-point modification impacts hydration at the anode inlet. Previous studies have shown that increasing the cathode inlet pressure and lowering the coolant outlet temperature raise the relative humidity of the anode gas, making the membrane more hydrated \cite{goshtasbi2021model,xu2022closed}. Similarly, a higher local current density near the anode inlet increases the production of ORR water, thus enhancing the membrane hydration in that region. Meanwhile, changes in the air flow rate have negligible impact on membrane hydration at the anode inlet \cite{goshtasbi2021model}. With this understanding, we next investigate how modifying the set-points can improve hydration when the stack experiences transient membrane dry-out at the anode inlet during a load reduction, as shown in Fig.~\ref{fig:hydration_transition}.

According to the set-points computation of the FCS in Section~\ref{sec:setpoint_map}, a load reduction is followed by a decrease in current, pressure, and temperature so that the FCS can deliver the requested power with maximum efficiency at steady-state. While the reduction in temperature improves membrane hydration at the anode inlet during dry-out, the reduction in current and tracking of the optimal pressure set-point result in a transient violation of the hydration constraint in this region. Since the fuel cell power output is directly governed by the current, modifying the current to enforce the hydration constraints would degrade the power tracking performance. Therefore, the motivation behind using the CG is to temporarily increase the pressure by adjusting its optimal set-point during dry-out, thereby maintaining membrane hydration at the anode inlet above its lower limit without modifying the current. As for the flow rate set-point, although it does not affect membrane hydration at the anode inlet, it must be adjusted in accordance with the modified pressure set-point to prevent potential compressor surge during load reduction.

To implement the hydration management CG, we begin by discretizing the reduced-order closed-loop linear model from Section~\ref{sec:Control-oriented model} using a sampling time of $T_s$ = 0.1 seconds. 
This model is derived around an equilibrium condition where the membrane hydration at the anode inlet is 6.144 [-]. As noted in Section~\ref{sec:Control-oriented model}, to prevent membrane dry-out, the hydration level at the anode inlet must remain above a lower limit, defined as, \( \lambda_{\text{mb,lb}} = 6 \). The discrete-time model, together with the hydration constraint, is then used to define the admissible set $\Omega$, as shown in Eq.~\eqref{eqn:omega}. Hydration constraint violations at all times can be avoided if $(x(k), v) \in \Omega$, where $x(k)$ is the internal state of the reduced-order model. Since the system states are not directly measurable, an observer is needed to estimate them based on available measurements such as EIS (see Fig.~\ref{fig:total_scheme}). However, the current model does not include the EIS dynamics, so we adopt a simplifying assumption in which membrane hydration at the anode inlet is treated as a measured output and used in a Kalman filter for state estimation. Although observer design for distributed hydration using physical measurements is an important challenge, it is beyond the scope of this work. However, this assumption allows us to focus on controlling the membrane hydration, which is the focus of this study. 

For real-time implementation of the CG in localized hydration management, solving the QP over the full admissibility index \( k^* \) can be prohibitive because of the high computational burden. In addition, applying the linear CG to a nonlinear system may lead to infeasibility of the online QP due to plant-model mismatch. This infeasibility can cause a governor ``hang-up'', wherein the governor stops updating the command. To address both of these issues in real-time implementation, we adopt two modifications: (1) a shorter prediction horizon \( p \ll k^* \) is used when computing the matrices \( H_x \), \( H_v \), and \( h \) in Section~\ref{sec:CG_review}; and (2) the linear inequality constraints in \( \Omega \) are softened by introducing a slack variable \( \varepsilon \), such that all constraints are imposed as soft constraints. These modifications eliminate the feasibility issue and make the QP computationally tractable for real-time implementation. With these modifications, the following QP is solved at each time instant \( k \) to adjust the optimal pressure set-point, if necessary:
\begin{equation}
\begin{aligned}
\big(p_{\mathrm{sm,ca}}^{\mathrm{mod}}(k),\,\varepsilon(k)\big)
&= \arg\min_{p_{\mathrm{sm,ca}}^{\mathrm{mod}},\,\varepsilon}
   \ \|p_{\mathrm{sm,ca}}^{\mathrm{mod}} - p_{\mathrm{sm,ca}}^{\mathrm{des}}(k)\|^2
   + \rho\, \varepsilon^2 \\
\text{s.t.}\quad
& -H_x \hat{x}(k) - H_v v \le h + \varepsilon\, \mathbf{1}, \\
& \varepsilon \ge 0 .
\end{aligned}
\label{eqn:hydration_qp}
\end{equation}
where \( v = [I_{\mathrm{d}}(k),\ T_{\mathrm{c,in}}^{\mathrm{des}}(k),\ T_{\mathrm{c,out}}^{\mathrm{des}}(k),\ W_{\mathrm{cp}}^{\mathrm{des}}(k),\ p_{\mathrm{sm,ca}}^{\mathrm{mod}}]^\top \), \( h \) is defined as before with \( s = 0.144 \), \( \rho \) is a slack penalty weight chosen to be large to strongly penalize constraint violations, and \( \mathbf{1} \) denotes a vector of ones with appropriate dimension.

As explained previously, when the pressure set-point is modified to $p_{\mathrm{sm,ca}}^{\mathrm{mod}}$, the flow rate set-point must also be adjusted to prevent the compressor trajectory from violating the surge constraint. While it is possible to incorporate the surge constraint into the CG formulation in \eqref{eqn:hydration_qp}, doing so introduces an additional decision variable and increases the computational burden of real-time implementation. To maintain the simplicity of the QP problem in \eqref{eqn:hydration_qp} and still address the surge constraint, we adopt a heuristic approach that mitigates the surge risk by using a conservative surge boundary, instead of explicitly incorporating the air supply dynamics into the CG design, as is done in~\cite{vahidi2006constraint}.
\begin{figure}[t!]
    \centering
    \includegraphics[width=0.48\textwidth]{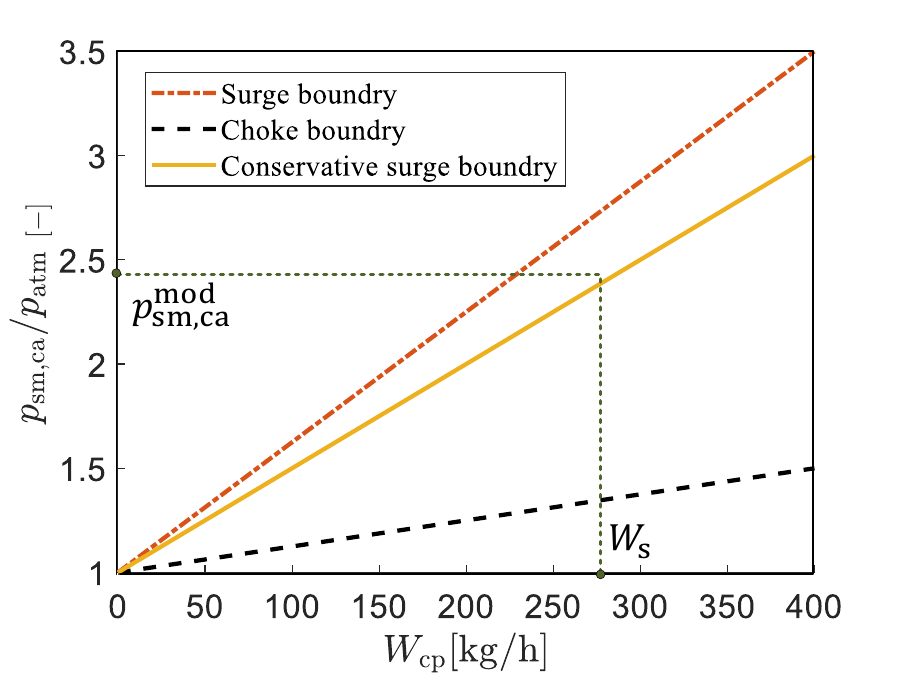}
    \caption{Surge, choke, and conservative surge boundaries in the compressor operating map. The surge region lies to the left of the dashed red line, while the choke region lies to the right of the dashed black line. The standard surge boundary is defined by $W_{\mathrm{cp}} = 160 \left( p_{\mathrm{sm,ca}} / p_{\mathrm{atm}} - 1 \right)$, the conservative surge boundary by \mbox{$W_{\mathrm{cp}} = 200 \left( p_{\mathrm{sm,ca}} / p_{\mathrm{atm}} - 1 \right)$}, and the choke boundary by $W_{\mathrm{cp}} = 800 \left( p_{\mathrm{sm,ca}} / p_{\mathrm{atm}} - 1 \right)$. Note that the plotted surge and choke boundaries serve only to illustrate the strategy's effectiveness and should not be interpreted as the actual compressor limits.
}
    \label{fig:conservative_surge}
\end{figure}

To implement this, the conservative surge boundary shown in Fig.~\ref{fig:conservative_surge} is used to determine the minimum admissible flow rate $W_s$ for a given \( p_{\mathrm{sm,ca}}^{\mathrm{mod}} \). If the optimal flow rate set-point \( W_{\mathrm{cp}}^{\mathrm{des}}(k) \) exceeds the minimum admissible value \( W_s(k) \), we set \( W_{\mathrm{cp}}^{\mathrm{mod}}(k) = W_{\mathrm{cp}}^{\mathrm{des}}(k) \), indicating that the operating point \( (p_{\mathrm{sm,ca}}^{\mathrm{mod}}, W_{\mathrm{cp}}^{\mathrm{mod}}) \) lies to the right of the conservative surge boundary, within the presumed safe region. Otherwise, we clip the flow rate to \( W_{\mathrm{cp}}^{\mathrm{mod}}(k) = W_s(k) \) to avoid operating in the surge-prone region.

In the following subsections, we evaluate the capability of the proposed CG strategy to manage membrane hydration at the anode inlet through two case studies. The first case study considers an FCS undergoing a current demand reduction following high-temperature operation, as illustrated in Fig.~\ref{fig:hydration_transition}. The second case study investigates the utility of the proposed CG framework when tracking a time-varying power demand.


\subsubsection{Application of the hydration management CG to a current demand reduction}
\begin{figure}[t!]
\centering
\subfloat[]{\includegraphics[width=0.33\textwidth]{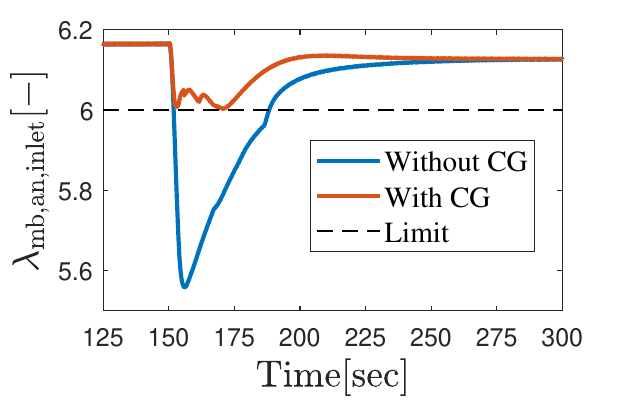}}\hspace{-0.5em}
\subfloat[]{\includegraphics[width=0.33\textwidth]{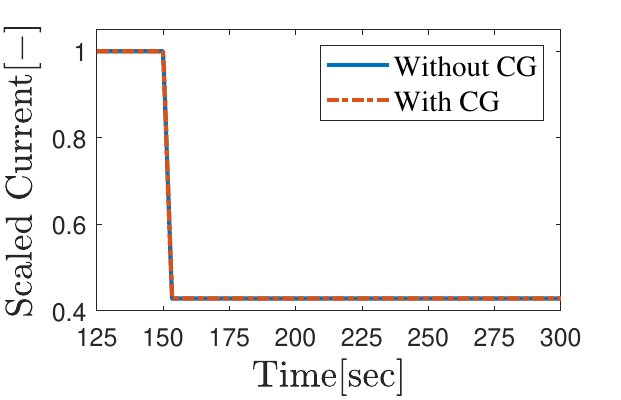}}\hspace{-0.5em}
\subfloat[]{\includegraphics[width=0.33\textwidth]{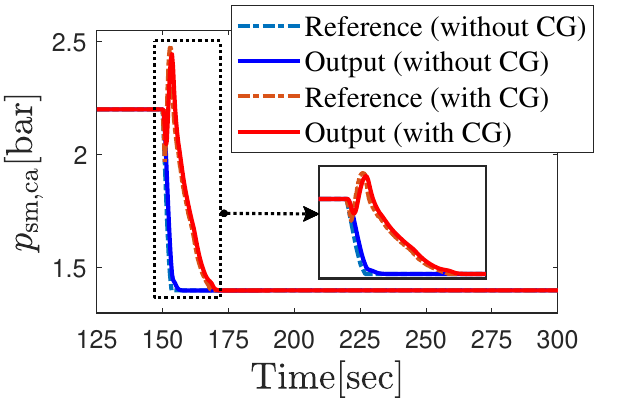}}\\[-1em]
\subfloat[]{\includegraphics[width=0.33\textwidth]{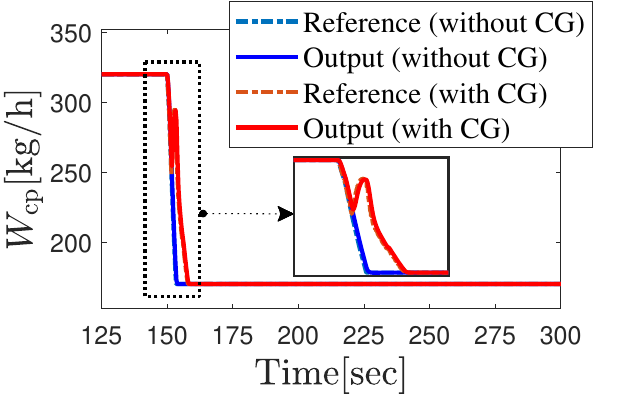}}\hspace{-0.5em}
\subfloat[]{\includegraphics[width=0.33\textwidth]{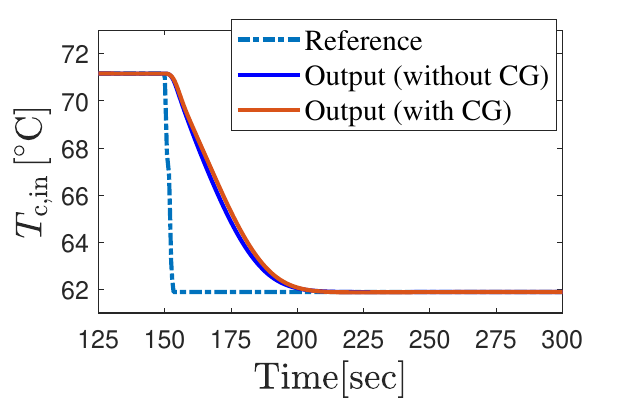}}\hspace{-0.5em}
\subfloat[]{\includegraphics[width=0.33\textwidth]{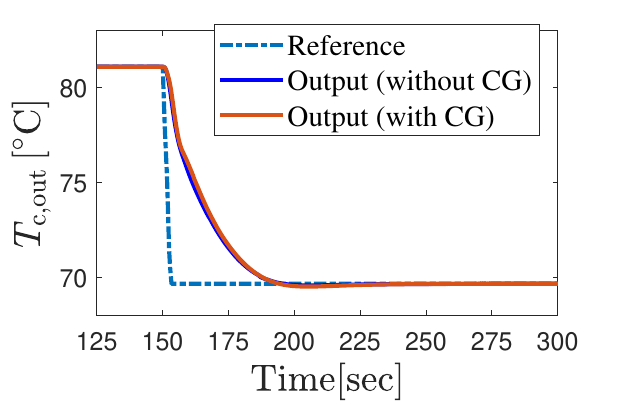}}\\[-1em]
\subfloat[]{\includegraphics[width=0.33\textwidth]{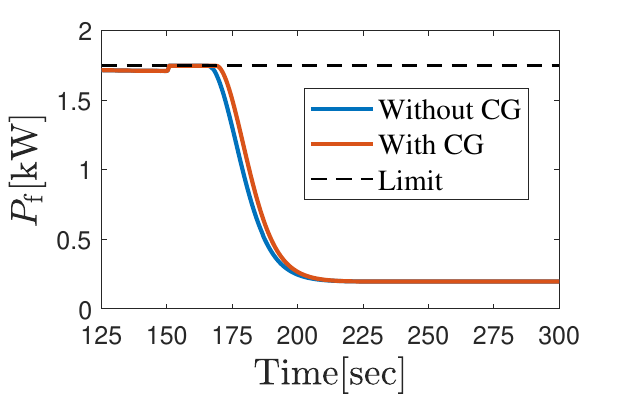}}\hspace{-0.5em}
\subfloat[]{\includegraphics[width=0.33\textwidth]{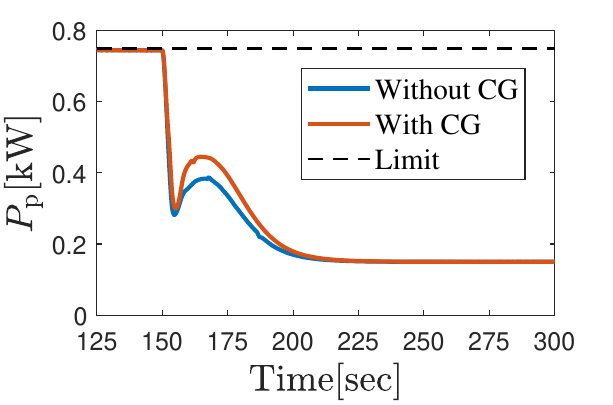}}\hspace{-0.5em}
\subfloat[]{\includegraphics[width=0.33\textwidth]{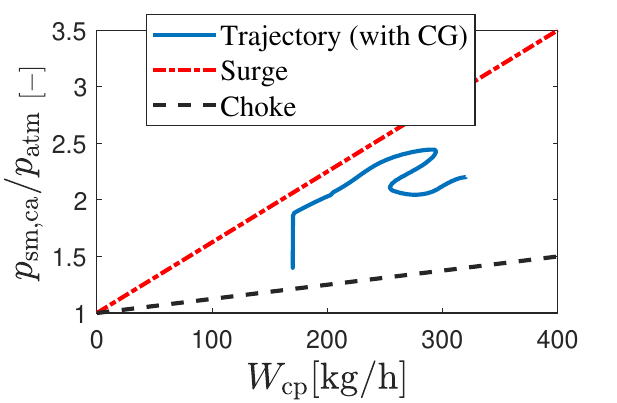}}
\caption{CG-based constraint management results: System state trajectories during a current demand reduction following high-temperature operation. Subplots depict: (a) membrane hydration at the anode inlet,(b) scaled stack current, (c) cathode supply manifold pressure, (d) compressor air mass flow rate, (e) coolant inlet temperature, (f) coolant outlet temperature, (g) fan power consumption, (h) pump power consumption, and (i) compressor response. }
\label{fig:CG_current_reduction}
\end{figure}

Simulation results of the high-fidelity FCS model subjected to a current reduction following high-temperature operation, both without and with the proposed CG-based constraint management strategy, are shown in Fig.~\ref{fig:CG_current_reduction}.
The results indicate that the proposed hydration management CG maintains membrane hydration at the anode inlet above its lower limit without any constraint violations (see Fig.~\ref{fig:CG_current_reduction}(a)), and does so without modifying the current demand, as shown in Fig.~\ref{fig:CG_current_reduction}(b). This is made possible by allowing the CG to temporarily raise the pressure set-point, starting at $t = 151$ seconds when the linear model predicts a potential violation of the hydration constraint, and continuing until $t = 153.5$ seconds, when there is no longer a potential risk of hydration violation due
to current reduction. After this point, the cathode pressure resumes tracking its set-point as quickly as possible, without compromising the hydration constraint (see Fig.~\ref{fig:CG_current_reduction}(c)). The flow rate set-point shown in Fig.~\ref{fig:CG_current_reduction}(d)) is also modified to avoid violation of the surge constraint, as illustrated in Fig.~\ref{fig:CG_current_reduction}(i)). As for the cooling system, the coolant inlet and outlet temperatures remain nearly the same in both the cases without and with the hydration management CG, as shown in Fig.~\ref{fig:CG_current_reduction}(e)) and (f). This is because, the controller does not apply a higher current to enforce the hydration constraint, and instead manages membrane hydration using only the air supply system. Another noteworthy observation is the saturation of the cooling system actuators during current reduction, as shown in Figs.~\ref{fig:CG_current_reduction}(g) and (h), which highlights the critical role of the air supply system in water management when the cooling system is unable to reduce the temperature due to actuator limitations.

\subsubsection{Application of the hydration management CG to a time-varying power demand}
\begin{figure}[b!]
\centering
\subfloat{%
  \includegraphics[width=1\textwidth]{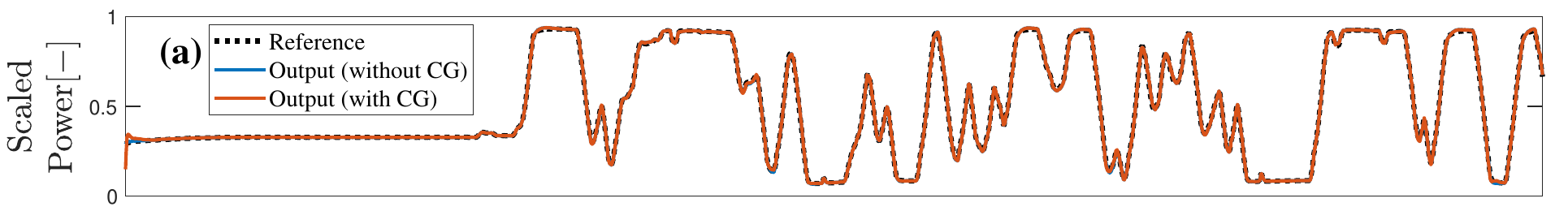}
}\\[-0.5em]
\subfloat{%
  \includegraphics[width=1\textwidth]{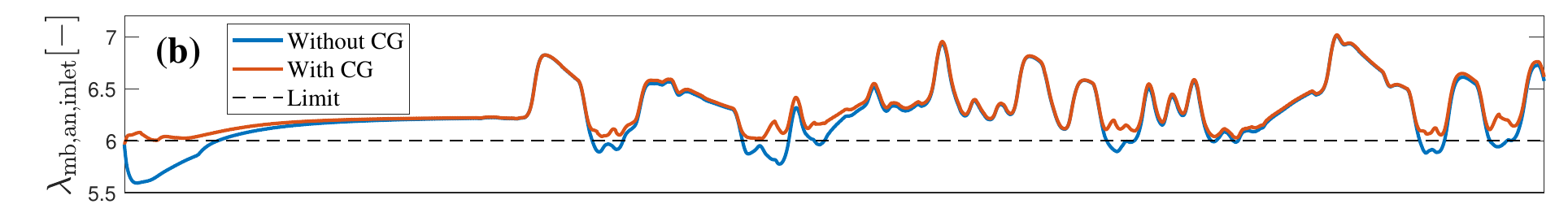}
}\\[-0.4em]
\subfloat{%
  \includegraphics[width=1\textwidth]{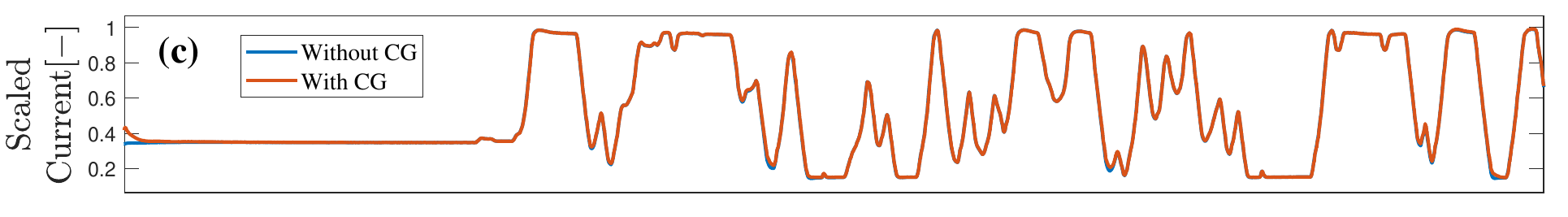}
}\\[-0.5em]
\subfloat{%
  \includegraphics[width=1\textwidth]{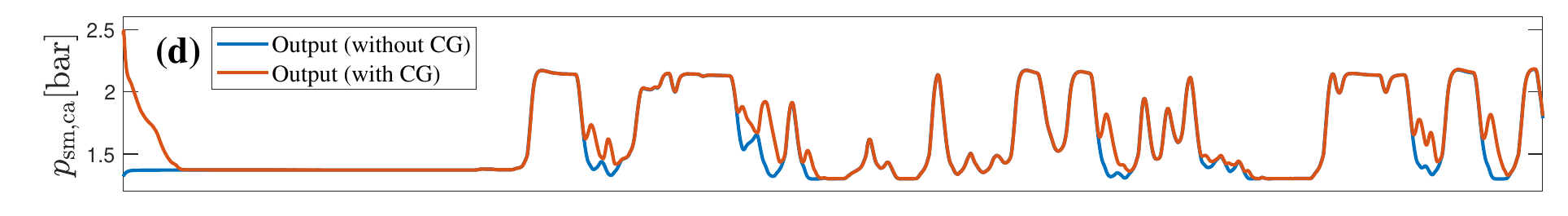}
}\\[-0.5em]
\subfloat{%
  \includegraphics[width=1\textwidth]{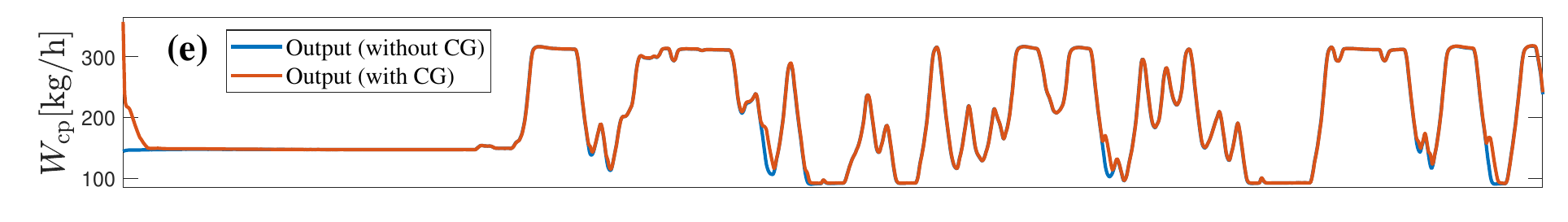}
}\\[-0.5em]
\subfloat{%
  \includegraphics[width=1\textwidth]{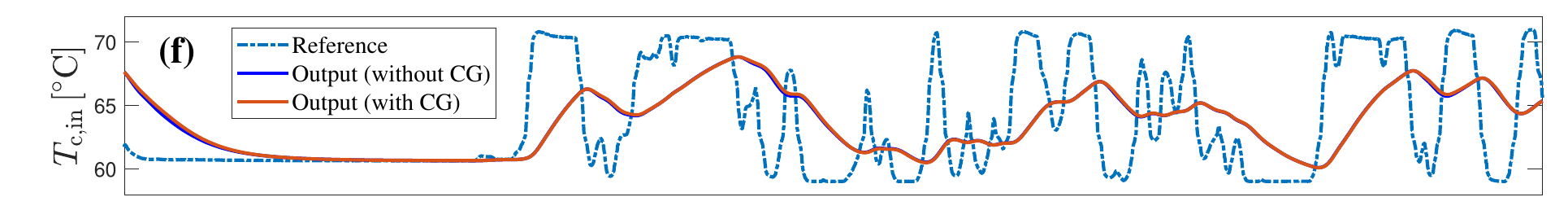}
}\\[-0.5em]
\subfloat{%
  \includegraphics[width=1\textwidth]{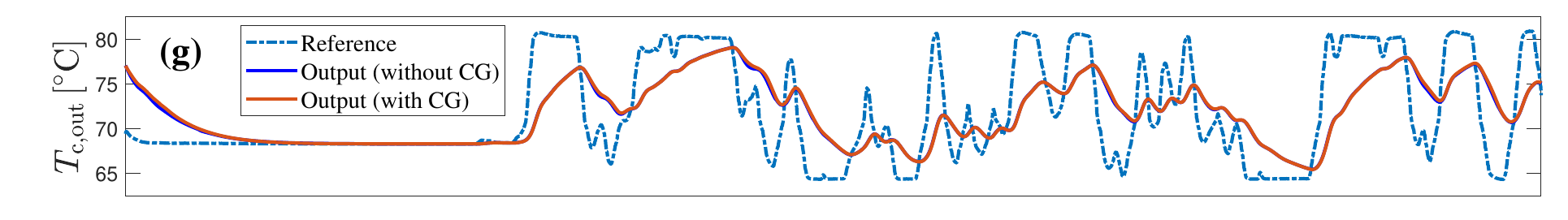}
}\\[-0.5em]
\subfloat{%
  \includegraphics[width=1\textwidth]{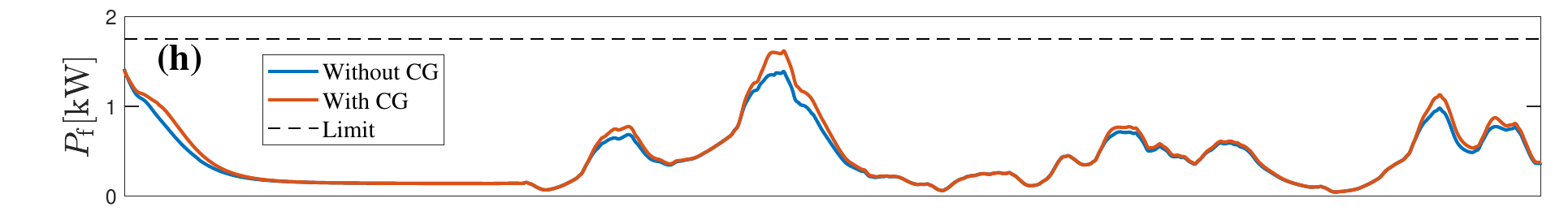}
}\\[-0.5em]
\subfloat{%
  \includegraphics[width=1\textwidth]{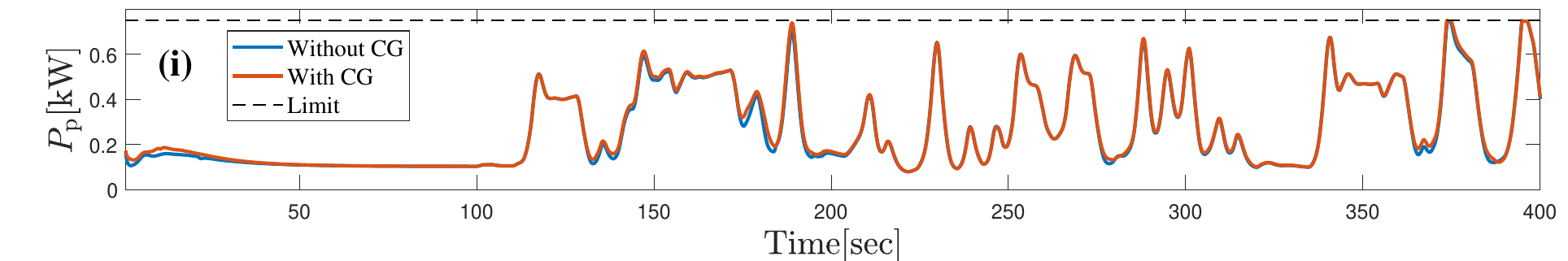}
}
\caption{CG-based constraint management results: Power, current, and system state trajectories during the “Dynamic Test” drive cycle. Subplots show: (a) scaled net power output, (b) membrane hydration at the anode inlet, (c) scaled stack current, (d) cathode supply manifold pressure, (e) compressor air mass flow rate, (f) coolant inlet temperature, (g) coolant outlet temperature, (h) fan power consumption, and (i) pump power consumption. The air supply set-points are not shown, as they are perfectly tracked by the LQI controller, as previously demonstrated in Fig.~\ref{fig:CG_current_reduction}}
\label{fig:CG_drive_cycle}
\end{figure}
In this section, the power tracking capability of the proposed CG strategy is evaluated in response to a time-varying power demand over a standard drive cycle, referred to as the ``Dynamic Test''. This test represents a more realistic scenario for automotive applications, where constraint satisfaction must be ensured throughout the operation. For power tracking, here the PI gains in Eq.~\eqref{eqn:Power_PI} are selected as $K_p = 0.5$ and $K_i = 0.05$. 

\begin{figure}[t!]
\centering
\includegraphics[width=0.48\textwidth]{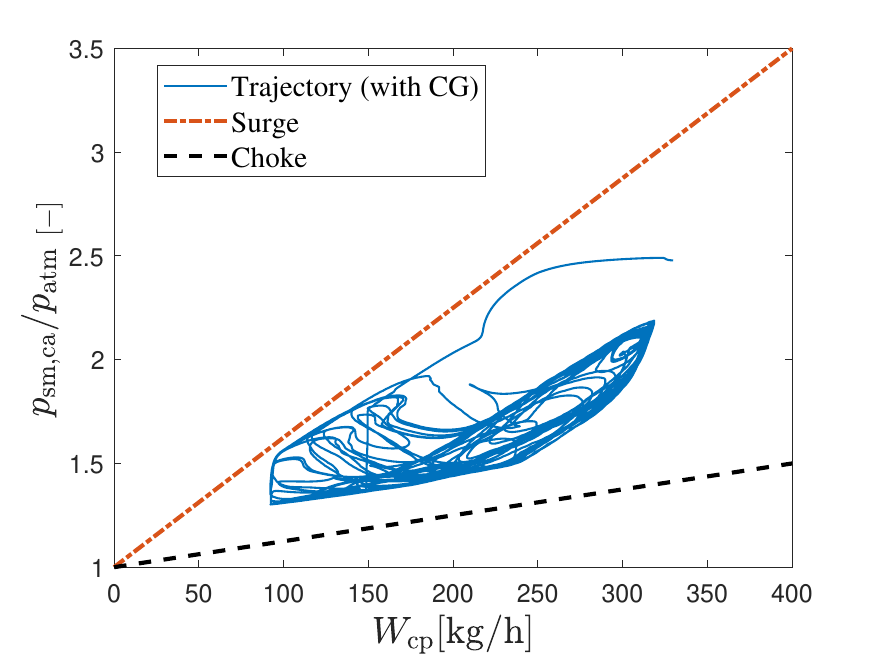}
\caption{Compressor operating trajectory during the “Dynamic Test” drive cycle under CG-based constraint management.}
\label{fig:CG_compressor_trajectory}
\end{figure}

Figure~\ref{fig:CG_drive_cycle} shows the power demand profile alongside the net power generated by the FCS. During current reductions, when the controller predicts a violation of the membrane hydration constraint at the anode inlet, it increases the cathode pressure accordingly. Once the hydration conditions begin to improve, due to increased current and/or reduced temperature, the pressure resumes tracking its optimal set-point as quickly as possible without compromising constraint satisfaction. Additionally, modifying the flow rate set-point in accordance with changes in the pressure set-point confines compressor operation between the surge and choke lines, as shown in Fig.~\ref{fig:CG_compressor_trajectory}. Thus, the proposed CG-based constraint management enables the power tracking controller to respond quickly to the power demand while ensuring that the hydration constraints are satisfied. It should be noted that the small difference observed in the current profiles in Fig.~\ref{fig:CG_drive_cycle}(c), with and without the CG, is not due to any CG action on the current itself. Rather, this difference results from the power tracking controller, which slightly adjusts the current to maintain accurate tracking of the net power reference in response to variations in the pressure and flow set-points used for hydration management and surge avoidance.

\section{Conclusions and outlook}
\label{sec:Conclusion}
In this article, a comprehensive nonlinear model of an automotive polymer electrolyte membrane (PEM) fuel cell system was employed to support the development of a constraint-aware control strategy for localized water management. The model incorporated a pseudo two-dimensional (P2D) representation of the stack and captured the interactions among key balance-of-plant components. To optimize the system’s efficiency, output feedback controllers were designed to track the optimal set-points of the air and coolant supply systems based on steady-state power optimization.

To enforce the membrane hydration constraint near the anode inlet, the driest region of the stack under counter-flow operation, a reduced-order linearized model of hydration dynamics was derived to capture its essential behavior with sufficient fidelity. This model enabled the implementation of a linear command governor (CG) as a computationally efficient constraint enforcement strategy. Given the slow dynamics of the thermal system, the air supply system, characterized by its faster response, was used for transient constraint satisfaction. Specifically, the CG adjusted the air supply pressure set-point during load reductions, when necessary, to maintain the membrane hydration above its lower limit without modifying the current demand.
The proposed control framework was evaluated through closed-loop simulations under realistic drive-cycle conditions. The simulation results demonstrated that the CG effectively mitigated hydration constraint violations while preserving power tracking performance.

Future work will explore improved methods for estimating the spatial distribution of membrane hydration to eliminate the simplified observer assumption adopted in this study. Additionally, the integration of short-horizon load preview information into the hydration management strategy will be investigated, as good-quality command previews are available in real-world scenarios and can allow the cooling subsystem to contribute more effectively to hydration control during transients.

\section*{Acknowledgments}
We acknowledge support from Ford Motor Company under award 002092-URP and U.S. Department of Energy (DOE) under award DE-EE0010147. The views expressed in this paper  do not necessarily
represent the views of the DOE, the United
States Government, or Ford Motor Company.

\section*{Appendix A}
This appendix presents the system matrices of the control-oriented model introduced in Section~\ref{sec:Control-oriented model}.

\begin{table}[h!]
\centering
\begin{tabular}{c|rrrrrrrr}
$A_{\text{cl}}$ & x1 & x2 & x3 & x4 & x5 & x6 & x7 & x8 \\
\hline
x1 & -0.5316 & -0.3919 & 0.1285 & -0.1022 & -1.8970 & 0.0065 & -0.3225 & -3.0660 \\
x2 & -0.0152 & -0.1734 & 0.1028 & 0.0819 & -0.7209 & 0.0937 & -0.3186 & -2.3680 \\
x3 &  0.0248 & -0.1797 & -0.0390 & -0.0399 & -1.2250 & -0.4064 & 0.8389  & 3.9420 \\
x4 & -0.3170 &  0.1394 & -0.0648 & -0.2977 & 1.0490  & 0.1525  & -0.2452 & -1.0040 \\
x5 &  1.2920 &  0.2207 & 0.2919  & -0.8066 & -3.5120 & -1.1330 & 2.7170  & 10.3400 \\
x6 &  0.0749 & -0.0586 & 0.0250  & 0.0366  & -0.7447 & -0.1458 & 0.2492  & 0.4318 \\
x7 & -0.4137 &  0.1608 & 0.0068  & 0.2372  & 2.5500  & 1.0500  & -2.4430 & -9.1920 \\
x8 & -0.8649 &  0.4470 & 0.4212  & 1.3470  & 6.9400  & 4.4760  & -10.580 & -48.590 \\
\end{tabular}
\end{table}
\begin{table}[h!]
\centering
\begin{tabular}{c|rrrrr}
$B_{v}$ & $I_{\text{d}}$ & $T_{\text{c,in}}^{\text{des}}$ & $T_{\text{c,out}}^{\text{des}}$ & $W_{\text{cp}}^{\text{mod}}$ & $p_{\text{sm,ca}}^{\text{mod}}$ \\
\hline
x1 & -0.00278 &  0.02772 & -0.00918 & -0.00145 & -0.66700 \\
x2 & -0.00179 &  0.03322 &  0.03176 & -0.00161 & -0.58070 \\
x3 & -0.00249 & -0.11090 &  0.00343 &  0.00510 & -0.37260 \\
x4 &  0.00241 &  0.01543 &  0.03885 & -0.00069 &  0.58880 \\
x5 & -0.00762 &  0.13350 &  0.03305 &  0.01283 & -0.22210 \\
x6 & -0.00606 & -0.00803 & -0.00307 &  0.00520 & -0.24330 \\
x7 & -0.00016 &  0.17820 &  0.05243 & -0.00831 &  0.57590 \\
x8 &  0.00298 &  0.75630 &  0.16840 & -0.04961 &  0.02221 \\
\end{tabular}
\end{table}
\vspace{-1.0em}
\begin{table}[h!]
\centering
\begin{tabular}{c|rrrrrrrr}
$C_{\text{cl}}$ & x1 & x2 & x3 & x4 & x5 & x6 & x7 & x8 \\
\hline
$\lambda_{\mathrm{mb},\mathrm{an,inlet}}$ & 0.5391 & -0.1317 & -0.0153 & 0.3863 & -0.0397 & -0.0290 & 0.0544 & 0.3306 \\
\end{tabular}
\end{table}
\vspace{-1.0em}
\begin{table}[h!]
\centering
\begin{tabular}{c|rrrrr}
$D_{v}$ & $I_{\text{d}}$ & $T_{\text{c,in}}^{\text{des}}$ & $T_{\text{c,out}}^{\text{des}}$ & $W_{\text{cp}}^{\text{mod}}$ & $p_{\text{sm,ca}}^{\text{mod}}$ \\
\hline
$\lambda_{\mathrm{mb},\mathrm{an,inlet}}$ & $-7.71 \times 10^{-6}$ & -0.004093 & -0.001124 & 0.000191 & 0.000827 \\
\end{tabular}
\end{table}

\setcounter{table}{0}
\renewcommand{\thetable}{A\arabic{table}}

{\footnotesize
\bibliographystyle{elsarticle-num}
\bibliography{BiblioTex}
\biboptions{sort&compress}}
\end{document}